\documentclass[useAMS,usenatbib]{mn2e}

\def\aap{{A\&A}}

\def\apj{{ApJ}}

\def\araa{{ARA\&A}}
\def\mnras{{MNRAS}}
\def\nat {{Nature}}

\usepackage{amsmath}
\usepackage{amssymb}
\usepackage{graphicx}
\title[Primordial IMF with Stellar Archaeology]{Constraining the primordial initial mass function with stellar archaeology}
\author[T.\ Hartwig, V.\ Bromm, R.\ S.\ Klessen, S.\ C.\ O.\ Glover]
{\parbox{\textwidth}{
Tilman Hartwig$^{1,2,3}$\thanks{E-mail: hartwig@iap.fr}, Volker Bromm$^{2}$, Ralf S.\ Klessen$^{1,4,5}$\\
and Simon C.\ O.\ Glover$^{1}$\\
}\\
$^{1}$Universit\"at Heidelberg, Zentrum f\"ur Astronomie, Institut f\"ur Theoretische Astrophysik, Albert-Ueberle-Str.\ 2, 69120 Heidelberg,\\
Germany\\
$^{2}$Department of Astronomy, University of Texas, Austin, Texas 78712, USA\\
$^{3}$Institut d'Astrophysique de Paris (UMR 7095: CNRS \& UPMC), 98bis Boulevard Arago, F-75014 Paris, France\\
$^{4}$Department of Astronomy and Astrophysics, University of California, 1156 High
Street, Santa Cruz, CA 95064, USA\\
$^{5}$Kavli Institute for Particle Astrophysics and Cosmology, Stanford University,
SLAC National Accelerator Laboratory, Menlo Park,\\
CA 94025, USA}

\begin{document}

%\date{Accepted date. Received date; in original form date}

\pagerange{\pageref{firstpage}--\pageref{lastpage}} \pubyear{2014}

\maketitle

\label{firstpage}

\begin{abstract}
We present a new near-field cosmological probe of the initial mass function (IMF)
of the first stars. Specifically, we constrain the lower-mass limit of the Population~III
(Pop~III) IMF with the total number of stars in large, unbiased surveys of the Milky Way. We model the early star formation history in a Milky Way-like halo with
a semi-analytic approach, based on Monte-Carlo sampling of dark matter merger trees, combined with a treatment of the most important feedback mechanisms. Assuming a logarithmically flat Pop~III IMF and varying its low mass limit, we derive the number of expected survivors of these first stars, using them to estimate the probability to detect any such Pop~III fossil in stellar archaeological surveys.
Following our analysis, the most promising region to find possible Pop~III survivors is the stellar halo of the Milky Way, which is the best target for future surveys.
We find that if no genuine Pop~III survivor is detected in a sample size of $4 \times 10^6$ ($2 \times 10^7$) halo stars with well-controlled selection effects, then we can exclude the hypothesis that the primordial IMF extended down below $0.8 M_\odot$ at a confidence level of $68\%$ ($99\%$).
With the sample size of the Hamburg/ESO survey, we can tentatively exclude Pop~III stars with masses below $0.65 M_\odot$ with a confidence level of $95\%$, although this is subject to significant uncertainties. To fully harness the potential of our approach, future large surveys are needed that employ uniform, unbiased selection strategies for high-resolution spectroscopic follow-up.
\end{abstract}


\section{Introduction}
The birth of the first, so called Population~III (Pop~III), stars marks the transition from the `Dark Ages' of the Universe to the complex structure we can observe today \citep{bromm13,loeb13,greif14}. These Pop~III stars synthesise the first heavy elements, thus enabling the formation of subsequent generations of stars, contribute to the early reionisation of the Universe, and might provide seeds for supermassive black holes \citep[for reviews, see][]{barkana07,volonteri12,kbbh13}. Although the first stars fundamentally influenced early cosmic evolution, there are so far no direct observations of them to guide theoretical understanding. A main goal of current research is to constrain the initial mass function (IMF) for Pop~III stars, because it is the key unknown in modelling their impact on cosmic history. Computer simulations predict that primordial star formation created stars with higher characteristic mass compared to the present day case. While early simulations suggested the formation of stars with 
masses 
above 
$100M_\odot$ \citep{op01,op03,abn02,bcl02}, more recent work shows that the accretion discs around Pop~III protostars
fragment, resulting in
multiple systems with masses between $10-100 M_\odot$ \citep{sgb10,clark11,getal11b,greif12,hyoy12,hetal14,hcgk14}. In extreme cases, this range might even extend down to $0.1 M_\odot$ \citep{dgck13,sb14}.

How can those theoretical predictions be empirically tested? In the absence
of any {\it in situ} detections of individual Pop~III stars, which will remain
largely out of reach even for the {\it James Webb Space Telescope (JWST)}, a promising 
alternative is stellar archaeology, the approach of scrutinising local fossils
for clues of the early Universe \citep[reviewed in][]{bc05,f10}. Specifically,
observations of elemental abundance patterns in extremely metal poor (EMP) stars allow us to discriminate individual supernova (SN) types that have enriched the gas out of which the next generation of (Pop~II) stars has formed. In principle, one has thus a handle on inferring the IMF of the first stars. Some constraints already exist. For example, current observations can be interpreted to limit the number of Pop~III stars that were massive enough ($>140 M_\odot$) to trigger hyper-energetic pair-instability supernovae (PISNe) to maximally $7\%$ of all Pop~III stars \citep{kjb08}. However, the actual mass range and the functional shape of the primordial IMF are still highly uncertain, as they remain elusive to direct empirical study. To make progress, we here propose a novel stellar archaeological test of the Pop~III IMF, targeting its crucial lower-mass limit.

The fact that no Pop~III survivor has been observed until now might suggest
that the lower limit to the primordial IMF exceeds $M_\mathrm{min}=0.8M_\odot$, such that all Pop~III stars would have died before reaching the age of the present-day Universe.
However, there are two key caveats which could greatly weaken this constraint.
First, any true low-mass Pop~III survivor could be `masqueraded' through
accretion of metal-enriched interstellar material from traversing the disc of the Milky Way, such that the survivor would appear as an extreme Pop~II star \citep{frebel09}. Recently, \citet{johnson14} studied the metal accretion from the ISM onto Pop~III stars and found that this process should lead to a unique chemical signature, because mainly gas phase elements can be accreted, while the radiation pressure prevents dust accretion. Consequently, it is possible that some of the so-called carbon-enhanced metal-poor stars could be polluted Pop~III stars, but this is still subject of an ongoing debate.
Second, and more seriously, it is not clear that existing surveys have sampled
a sufficient number of stars in either the Galactic halo or bulge to be sure that
no survivor went undetected. To address the second question, we model the detailed early assembly
history of the Milky Way, to obtain a realistic and statistically sound estimate of putative Pop~III low-mass
survivors. This in turn allows us to derive `critical' survey sample sizes that
need to be reached to effectively constrain the lower end of the primordial IMF.

We address this question by modelling the mass assembly history of a Milky Way-like halo with a semi-analytic merger tree approach, tracing the location of any low-mass Pop~III stars along the tree. Including all relevant feedback mechanisms, we derive the number of possible Pop~III survivors in the present-day Milky Way, together with their radial distribution.

Star formation crucially depends on the ability of the gas to cool in sufficiently short time. Since there are no metals present in the early Universe, the primordial gas cools mainly via H$_2$ line emission, when falling into a dark matter minihalo, predicted to be the formation site of the first stars \citep{haiman96a}. Some metal-free haloes, however, might have been affected by ionising radiation from neighbouring star-forming regions. The higher electron fraction left behind by this ionised regions could trigger the formation of hydrogen deuteride (HD), which serves as an additional coolant under these conditions, enabling the primordial gas to reach lower temperatures compared to those accessible with H$_2$ cooling only. Although several groups have analysed this second formation mode of primordial stars, there is no agreement how it might influence the characteristic stellar masses \citep{johnson06,yoshida07,cgkb11}.

Although several studies have already addressed the question whether to expect any Pop~III survivors in the Milky Way and where to look for them, this topic is still under considerable debate \citep{kbbh13}. Due to the inside-out growth of dark matter haloes, most studies predict that first star survivors should be concentrated towards the galactic centre \citep[e.g.][]{ws00,dmm05,bhp06,salvadori10,t10a,t10b}, while others propose they are spread over the entire Galaxy \citep{setal06,betal07}. Other studies, which use the same methodological approach, investigate the possibility that present-day galactic haloes might contain massive black holes (MBHs) which form by merging of black hole remnants of the first stars \citep{islam03,islam04a,islam04b}. These studies find that these MBHs will not be clustered towards the centre of the main halo, but rather continue to orbit within satellite subhalos. It is therefore not clear where to focus the search for Pop~III fossils, the Galactic bulge or extended 
halo.

There are several previous attempts to derive constraints on the primordial IMF based on stellar archaeology. \citet{t06} models the chemical evolution within the hierarchical build-up of the Milky Way, to investigate the contribution of the first stars to the chemical abundance record in low-metallicity stars. He finds that existing abundance constraints do not yet allow to distinguish between different Pop~III IMFs, but functions with characteristic mass of the order of a few $10 M_{\odot}$, compared to the previously preferred $100 M_{\odot}$, produce overall better fits to the available data. Similarly, \citet{salvadori07} study the stellar population history and chemical evolution of the Milky Way. By matching their predictions to the metallicity distribution function of metal-poor stars in the Galactic halo they find that Pop~III stars should be more massive than $0.9 M_{\odot}$. \citet{krhv13} explore the influence of Pop~III stars on the abundance patterns of damped Lyman-$\alpha$ absorbers
(DLAs), concluding that the DLA chemistry provides a sensitive probe of the primordial IMF, at least at sufficiently high redshifts. In a slightly different approach, \citet{mapelli06} derive an upper limit on the density of Galactic intermediate mass black holes (IMBHs), which
have been proposed in their model to be the relics of Pop~III stars. They compare the distribution of simulated X-ray sources with the observed one and base their conclusion on the null detection of any such source in the Galaxy. However, these IMBHs do not trace the low-mass end of the Pop~III IMF. In a recent study, \citet{debennassuti14} simulate the metallicity distribution function in the Galactic halo and compare this to stellar archaeology data. They find that faint SN explosions dominate the metal enrichment by the first stars, which in turn disfavours Pop~III stars in excess of $140 M_\odot$ and hence limits the upper mass end of the primordial IMF. However, none of these models has used the number of expected survivors together with the current sample sizes to directly constrain the lower mass limit of the primordial IMF.

Our paper has the following structure. In Section~2, we describe our methodology
to model the hierarchical assembly of the Milky Way. In Section~3, we compare our model to empirical constraints and we present the results in Section~4. In Section~5 we test the parameter sensitivity of our model by changing several basic assumptions. We summarise our results in Section~6.

\section{Methodology}
\begin{figure*}
\centering
\includegraphics[width=0.99\textwidth]{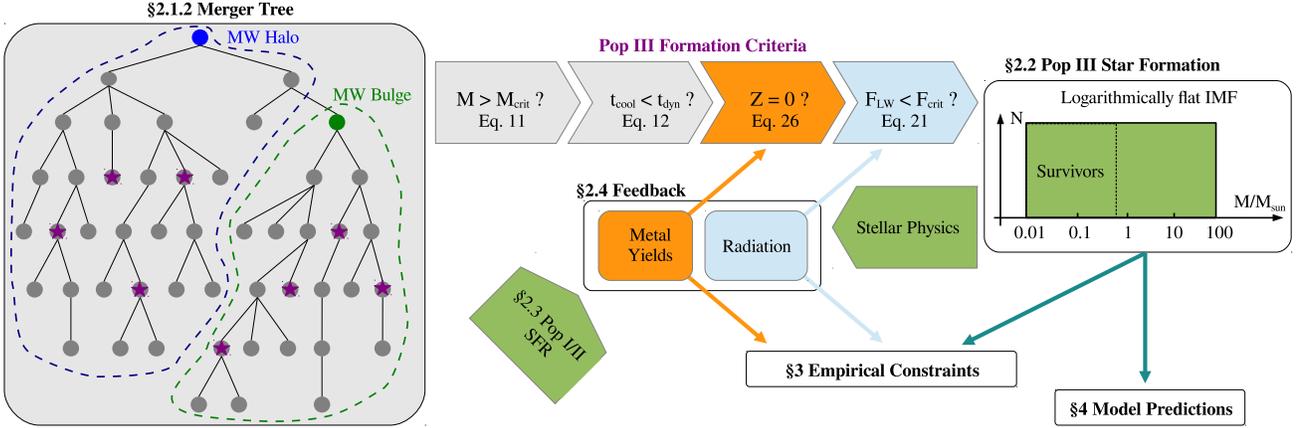}
\caption{Roadmap, illustrating our model, with references to the relevant sections and equations. Based on the merger tree, we check which haloes are able to form Pop~III stars. These checks include the critical mass, the absence of dynamical heating due to mergers, no pollution by metals and the strength of the LW background. We assign an individual number of Pop~III stars to each successful halo and determine the influence on their environment. The contribution of Pop I/II star formation is modelled based on the analytical cosmic star formation history. By comparing to existing observations, we can calibrate our model parameters. Finally, we derive a prediction for the number of Pop~III survivors in the Milky Way and determine constraints on the primordial IMF.}
\label{fig:modeloverview}
\end{figure*}

Here, we present our model of structure and star formation within the Milky Way, which is also illustrated in Fig.\;\ref{fig:modeloverview}. First, we discuss our implementation of hierarchical structure formation, present our model of the Milky Way, and describe our recipes for star formation and the related feedback mechanisms.

\subsection{Hierarchical Structure Formation}
Structure formation on cosmological scales is dominated by gravity. Tiny quantum fluctuations in the very early Universe imprinted density perturbations in the dark matter component, eventually leading to the collapse of overdensities. During cosmic evolution, small haloes successively merge together to form bigger and more massive objects, giving rise to the complex hierarchy of structure in the present-day Universe. In this section we present our methodology of determining the hierarchical mass assembly history of the Milky
Way.

We assume a flat $\Lambda$CDM Universe and use the cosmological parameters presented by the \citet{planck13} with additional constraints from \textit{WMAP} polarisation at low multipoles, high-resolution cosmic microwave background (CMB) data sets, and baryonic acoustic oscillations: $H_0=67.77 \mathrm{\;km\ s}^{-1} \mathrm{\;Mpc}^{-1}$, $\Omega_\Lambda = 0.6914$, $\Omega_m = 0.3105$, $\Omega_b = 0.04825$, $n_s=0.9611$, $\sigma _8 =0.8288$, $\tau=0.0952$, $Y_\mathrm{He} = 0.2477$. The dark matter power spectrum was calculated with the {\sc CAMB} code by \citet{lcl00} for wave numbers $10^{-6}\mathrm{\;Mpc}^{-1} \leq k/h \leq 10^{6}\mathrm{\;Mpc}^{-1}$, where $h$ is the Hubble constant in units of
$100\mathrm{\;km\ s}^{-1} \mathrm{\;Mpc}^{-1}$.

\subsubsection{Extended Press-Schechter Theory}
Originally, \citet{ps74} developed an analytical model to determine the mass assembly history of the Universe. Based on simple, generic assumptions, this `Press-Schechter' formalism is able to predict the number density of dark matter haloes as a function of their mass and redshift with surprising accuracy. The comoving number density of haloes of mass between $M$ and $M+\mathrm{d}M$ is given by
\begin{equation}
 \frac{\mathrm{d}n}{\mathrm{d}M} = \sqrt{\frac{2}{\pi}} \frac{\rho _m}{M} \frac{-\mathrm{d}(\ln \sigma (M))}{\mathrm{d} M} \nu _c \exp (- \nu _c ^2 /2)
\end{equation}
where the current matter density is $\rho _m$, the standard deviation of the matter power spectrum $\sigma (M)$, and $\nu _c = \delta _c (z)/ \sigma (M)$ the critical threshold for collapse. The time evolution of this critical overdensity for non-linear collapse is given by
\begin{equation}
 \delta _c (z) = \frac{1.686}{D(z)}
  \label{equ:deltac}
\end{equation}
with the linear growth factor being $D(z)$, normalised such that $D(0)=1$. \citet{bcek91} and \citet{lc93} improved this idea by interpreting the merger history of dark matter as a random walk in $k$-space, where $k$ is the wave number associated with density perturbations, smoothed on a scale $2 \pi / k$. This new idea also allows for the determination of merger rates and specific merger histories for individual objects.\\
We apply this extended Press-Schechter (EPS) approach to model more accurately the merger history of our Galaxy \citep{l10}. The probability distribution that a mass element finds itself at $z_2$ in a halo of mass $M_2$ that was at an earlier redshift $z_1$ part of a halo with mass $M_1 < M_2$ is given by the conditional probability
\begin{equation}
\begin{split}
\frac{\mathrm{d}P}{\mathrm{d}M_1} (M_1,z_1|M_2,z_2) = \sqrt{\frac{2}{\pi}} \frac{\delta _c (z_1) - \delta _c (z_2)}{[\sigma ^2 (M_1) - \sigma ^2 (M_2)]^{3/2}}\\
 \times \left| \frac{\mathrm{d} \sigma (M_1)}{\mathrm{d}M_1} \right| \exp \left( - \frac{[\delta _c (z_1) - \delta _c (z_2)]^2}{2[\sigma ^2 (M_1) - \sigma ^2 (M_2)]} \right).
\end{split}
 \label{eq:EPS}
\end{equation}
Hence, the probability that a halo at redshift $z_1$ is above a mass $M_0$ (and will end in a halo of mass $M_2$ at redshift $z_2$) is given by $P(>M_0,z_1|M_2,z_2)$ and the corresponding number density of haloes above the critical mass is given by
\begin{equation}
 n(>M_0) = \rho _m \int _{M_0} ^ \infty \frac{\mathrm{d}P(M_1,z_1|M_2,z_2)}{\mathrm{d}M_1} \frac{\mathrm{d}M_1}{M_1}.
\end{equation}
The original Press-Schechter formalism generally underestimates the number of low-mass haloes at high redshifts by almost an order of magnitude, compared with the results from cosmological simulations \citep{setal05,gb06,sasaki14}. However, we are explicitly interested in the exact number of these objects at high redshift. To overcome this shortcoming, we describe the mass assembly of the Milky Way with the dark matter halo merger tree algorithm by \citet{pch08}, which is originally based on the {\sc galform} package \citep{clbf00}. This code constructs merger trees, following the EPS formalism \citep{bcek91}, which reproduce the halo merger histories of the Millennium simulation \citep{setal05}.

\subsubsection{Merger Tree}
Here, we briefly describe the original code, and discuss the latest version by \citet{pch08} used for our work. Based on the conditional probability of the EPS formalism (Eq.~\ref{eq:EPS}), one can determine the limit of $z_1 \rightarrow z_2$ and derive the instantaneous merger rate
\begin{equation}
\begin{split}
\left. \frac{\mathrm{d}f}{\mathrm{d}z_1} \right| _{z_1 = z_2} \mathrm{d} \ln M_1 \mathrm{d} z_1 = \\
\sqrt{\frac{2}{\pi}} \frac{\sigma _1 ^2}{(\sigma _1 ^2 - \sigma _2 ^2)^{3/2}} \frac{\mathrm{d} \delta _1}{\mathrm{d} z_1} \left| \frac{\mathrm{d} \ln \sigma _1}{\mathrm{d} \ln M_1} \right| \mathrm{d} \ln M_1 \mathrm{d} z_1,
\end{split}
\end{equation}
where $f$ represents the fraction of mass from haloes of mass $M_2$ at redshift $z_2$ that is contained in progenitor haloes of mass $M_1$ at an earlier redshift $z_1$ and $\delta _1 = \delta _c (z_1)$ is the critical overdensity (Eq.~\ref{equ:deltac}) at redshift $z_1$. Consequently, the mean number of haloes of mass $M_1$ into which a halo of mass $M_2$ splits when one takes a step $\mathrm{d} z_1$ up in redshift (and hence backwards in cosmic time) is given by
\begin{equation}
 \frac{\mathrm{d} N}{\mathrm{d} M_1} = \frac{1}{M_1} \frac{\mathrm{d} f}{\mathrm{d} z_1} \frac{M_2}{M_1} \mathrm{d} z_1 \quad (M_1 < M_2).
 \label{eq:dNdM}
\end{equation}
For a mass resolution limit of $M_\mathrm{res}$, the mean number of progenitors with masses $M_1$ in the interval $M_\mathrm{res} < M_1 < M_2 /2$ can be expressed as
\begin{equation}
 P=\int _{M_\mathrm{res}} ^{M_2/2} \frac{\mathrm{d} N}{\mathrm{d} M_1} \mathrm{d} M_1,
 \label{eq:P}
\end{equation}
and the fraction of mass of the final object in progenitors below the resolution limit is given by
\begin{equation}
 F= \int _{0} ^{M_\mathrm{res}} \frac{\mathrm{d} N}{\mathrm{d} M_1} \frac{M_1}{M_2} \mathrm{d} M_1.
 \label{eq:F}
\end{equation}
Note, that the quantities $P$ and $F$ are proportional to the redshift step $\mathrm{d}z_1$ (Eq.~\ref{eq:dNdM}). For a given target mass and redshift, the {\sc galform} algorithm generates a corresponding binary merger tree backwards in time by choosing a redshift step $\mathrm{d}z_1$, such that $P \ll 1$, to ensure that the halo is unlikely to have more than two progenitors at the earlier redshift $z + \mathrm{d}z$. Next, it generates a uniform random number $R$, in the interval 0 to 1. If $R > P$, then the main halo is not split at this step. We simply reduce its mass to $M_2 (1 - F)$ to account for mass accreted in unresolved haloes. Alternatively, if $R \leq P$, then we generate a random value of $M_1$ in the range $M_\mathrm{res} < M_1 < M_2 /2$,
consistent with the distribution given by Eq.~(\ref{eq:dNdM}), to produce two new haloes with masses $M_1$ and $M_2 (1 - F) - M_1$. The same process is repeated for each new halo at successive redshift steps to build up a complete tree, which is finally stored at a limited number of output redshifts, so that each halo can have multiple progenitors at these discretised output redshifts.\\
The original {\sc galform} code systematically underpredicts the mass of the most massive progenitors for higher redshifts. Hence, we use the updated version of the code by \citet{pch08}, \textrm{which} modifies the progenitor mass function with a perturbing function
\begin{equation}
 \frac{\mathrm{d}N}{\mathrm{d} M_1} \rightarrow \frac{\mathrm{d}N}{\mathrm{d} M_1} G(\sigma _1 / \sigma _2 , \delta _2 / \sigma _2)
\end{equation}
to match the halo merger histories of the Millennium simulation \citep{setal05}. The best-fitting perturbing function is given by
\begin{equation}
 G(\sigma _1 / \sigma _2 , \delta _2 / \sigma _2) = 0.57 \left( \frac{\sigma _1}{\sigma _2} \right) ^{0.38} \left( \frac{\delta _2}{\sigma _2} \right) ^{-0.01}.
\end{equation}
We have chosen this specific implementation of the merger tree, because on the one hand it provides a fast algorithm to produce merger trees with arbitrary mass resolution and on the other hand, it performs best compared to other codes. \citet{jb14} recently compared four different implementations of merger trees and find the algorithm of \citet{pch08} to be the only one that yields the mass assembly history, merger rates, and the unresolved subhalo mass function in good agreement with simulations.\\

\subsubsection{Critical Mass for Baryonic Collapse}
Whether the primordial gas in a halo can collapse and form stars mainly depends on its ability to cool, which in turn depends on the abundance of molecular hydrogen in the early Universe. \citet{g13} models the H$_2$ abundance within low density gas falling into a dark matter minihalo and derives a formula for the critical halo mass by requiring that the gas must be able to cool in less than $20\%$ of the Hubble time. Only haloes above
\begin{equation}
 M_\mathrm{crit} = 6 \times 10^5 h^{-1} \left( \frac{\mu}{1.2} \right) ^{-3/2} \Omega _m ^{-1/2} \left( \frac{1+z}{10} \right) ^{-3/2} M_\odot
 \label{eq:mcrit}
\end{equation}
fulfil this criterion, where $\mu = 1.23$ is the mean molecular weight of neutral primordial gas. The cooling condition above is only a first order approximation because gas can also be heated during halo mergers \citep{yahs03}. Hence, dynamical heating from mass accretion and mergers opposes the relatively inefficient cooling by molecular hydrogen and therefore delays the formation of rapidly growing haloes \citep{wr78}. \citet{yahs03} identified the formation sites of Pop~III stars and included the effect of dynamical heating. They find that molecular cooling is more efficient than the dynamical heating only if a halo with mass $M$ has an instantaneous mass growth rate below
\begin{equation}
 \frac{\Delta M}{\Delta z} \lesssim 3.3 \times 10^6 M_\odot \left( \frac{M}{10^6 M_\odot} \right)^{3.9}.
 \label{eq:mdyn}
\end{equation}
Otherwise, the collapse is suppressed, or at least delayed, until the halo is massive enough to compensate this effect. We apply this criterion to our merger tree by checking for each halo that is above the critical mass whether it also fulfils this additional condition. The mass growth rate is the mass difference between the halo of interest and its most massive progenitor at the previous redshift step.

\subsubsection{Milky Way Characteristics}
Due to the inside-out growth of dark matter, we expect the Galactic bulge to contain the oldest stellar relics of the Milky Way. The disc has formed later and is not relevant for our stellar archaeology approach, because there are no stars in the disc with $[\mathrm{Fe}/\mathrm{H}] < -2.2$ \citep{fbh02}. The ancient thick disk, which might also contain very metal poor stars, is treated as a part of the stellar halo in our study. Moreover, we do not account for dwarf satellite galaxies or globular clusters here, because we do not have the required spatial information in our merger tree model. Phrased differently, we consider metal-poor dwarf galaxies as separate entities and we focus on the smooth stellar population, which also excludes globular clusters. Although some halo stars might have been contributed from disrupted globular clusters, this can only be a small contribution to the total stellar content.

At the high redshifts of interest, the formation sites of the first stars were homogeneously distributed within the comoving volume of the Milky Way, and some of their remnants may still be in the halo. Hence, we model the merger history of the Galactic halo and the early 
stellar bulge. For our model we have to define a `target' mass and redshift to create the corresponding mass assembly histories, where mass here represents the total mass of the halo or bulge, respectively, and the redshift marks the moment of virialisation of those components.
The Milky Way dark matter halo mass can be approximated by $M_\mathrm{halo}=(1.26 \pm 0.24) \times 10^{12} M_\odot$ \citep{m11}, which yields a comoving volume of $V_\mathrm{com,halo}=30.3 \mathrm{\;Mpc}^3$. The redshift however is not clearly defined, because virialisation is a gradual process. Assuming mass conservation, a physical radius of $100$\,kpc for the stellar halo \citep{fbh02}, and an overdensity at virialisation of $\Delta = 18 \pi ^2$ with respect to the mean cosmic density, we can approximate the redshift of virialisation by
\begin{equation}
 R_\mathrm{phys} ^3 \Delta (1+z_\mathrm{vir,halo})^3 = R_\mathrm{com} ^3,
\end{equation}
which yields $z_\mathrm{vir,halo}=2.5$. At this redshift, we assume the mass accretion history of the Milky Way to end. For the bulge mass we use the value of $M_\mathrm{bulge} = 1.8 \times 10^{10} M_\odot$ \citep{b95}, which is in good agreement with other determinations, e.g. by \citet{wd05}. Consequently, the comoving volume of the Galactic bulge is given by $V_\mathrm{com,bulge}=V_\mathrm{com,halo} M_\mathrm{bulge} / M_\mathrm{halo} = 0.43 \mathrm{\;Mpc}^3$. Note that we assume the same cosmic background density for both components. Although the stellar density differs in bulge and halo, this mass-weighted volume is a reasonable distinction between the two regions at higher redshifts. In any case, we only need these volumina to use and determine the star formation rates per comoving volume. Moreover, our main conclusions of the analysis are rather insensitive to the choice of the
bulge mass, since the Pop~III survivors in the halo outnumber those in the bulge by more than an order of magnitude.

For the chemical feedback model we need to know the physical volume of the Milky Way at any redshift, which can be determined by means of the spherical collapse model. At very high redshifts, the physical volume of the Milky Way expands with the Hubble flow until it decouples from it and undergoes gravitational collapse. The time evolution of the radius can be approximated in terms of the development angle $\theta$ by
\begin{equation}
\begin{split}
r(\theta) &= A(1- \cos \theta) \\
t(\theta) &= B(\theta - \sin \theta),
\end{split}
\end{equation}
where $A$ and $B$ are normalisation parameters. For this model, virialisation occurs at $\theta _\mathrm{vir} = 3 \pi /2$ with $r_\mathrm{vir} = r_\mathrm{max}/2 = A$. Consequently, the second free parameter is given by
\begin{equation}
 B = \frac{t_\mathrm{vir}}{3 \pi /2 + 1},
\end{equation}
where $t_\mathrm{vir}$ is the time of virialisation. We assume the virial radius of the Milky Way to be the same as its current physical radius. This approximation is valid within a factor of a few and yields reasonable results for our chemical feedback model (see Section~2.4.1).

For our modelling of the bulge, we assume that the first halo that has a mass of $M_\mathrm{bulge}$ will finally become the bulge, and that all Pop~III stars already present in this halo will end up in the present-day bulge. The redshift at which the first halo with $M_\mathrm{halo} > M_\mathrm{bulge}$ virialises is
\begin{equation}
 z_{\rm vir,bulge} = 9.1 \pm 0.5
\end{equation}
which is quite insensitive to changes in the bulge mass.
A distinction between bulge and halo based on this criterion is formally only valid as long as there are no major mergers or tidal stripping events. We have checked that the last merger of haloes, with a mass ratio of 1:3 or larger, occurs before $z=35$. However, this high redshift is a consequence of our small time steps ($\Delta z =0.16$) and should be interpreted with caution, because e.g. the ancient thick disk could have been a merger at smaller redshifts at about this ratio.

To test this approximation, we have compared our mass-dependent definition of the bulge to the results of a three-dimensional high-resolution simulation. Using the same criterion for the bulge and tracing its most bound particles to their current positions reveals that the bulge does not end up as centrally concentrated as expected (B. Griffen, priv. comm.). The final density of particles that trace the bulge has the same radial slope as the other dark matter particles. Phrased differently, most particles of the first halo with the bulge mass will not end up as the final bulge, which hence weakens our simple distinction criterion. However, this specific question on the spatial distribution of possible Pop~III survivors will be subject of a subsequent study and does not affect our final conclusion.

\subsection{Primordial IMF}
The overall mass range of primordial star formation is not yet well-known and subject to ongoing debate.
\citet{nu02} propose a bimodal IMF with a low-mass and a high-mass star formation mode, whereas simulations of Pop~III star formation predict a flat distribution of stellar masses. The possibility of disc fragmentation can lead to masses below $1M_\odot$ \citep[e.g.][]{dgck13,clark11}. Here, we assume a logarithmically flat IMF \citep{getal11b}
\begin{equation}
 \frac{\mathrm{d} N}{\mathrm{d} \ln M} = \mathrm{const} %\mbox{\ ,}
 \label{eq:IMF}
\end{equation}
and explore a mass range from
\begin{equation}
 M_\mathrm{min}=0.01 M_\odot
 \label{eq:Mmin}
\end{equation}
to
\begin{equation}
 M_\mathrm{max}=100 M_\odot ,
 \label{eq:Mmax}
\end{equation}
where the lower limit is close to the opacity-limit for fragmentation \citep{rees76}, and the upper limit is suggested by current simulations \citep{sgb12,hetal14}. The number of survivors is determined based on the mass, corresponding lifetime and the redshift of formation of the individual stars. Generally, Pop~III stars with masses below $0.8 M_\odot$ might survive until today \citep{mgcw01}, although there are individual possible survivors with masses up to $0.83 M_\odot$, which form at smaller redshifts.

The other key question is the amount of gas that ends up in stars per minihalo. Given the total mass of the minihalo $M_\mathrm{halo}$, the mass that ends in stars is given by
\begin{equation}
 M_* = \eta _* f_\mathrm{LW} \frac{\Omega _b}{\Omega _m} M_ \mathrm{halo},
\label{eq:etas}
\end{equation}
where the efficiency factor $f_\mathrm{LW}$ describes which fraction of the gas is able to collapse to cold and dense clouds under a given Lyman-Werner (LW) background and $\eta _*$ defines the fraction of this gas that will finally end in stars. Note that $\eta _*$ might itself depend on the halo mass, but for simplicity, we use a mean value. The fraction of cold, dense gas per minihalo under the influence of a LW-background is given by \citep{mba01}
\begin{equation}
 f_\mathrm{LW}= 0.06 \ln \left( \frac{M_\mathrm{halo} / M_\odot}{1.25 \times 10^5 + 8.7 \times 10^5 F_\mathrm{LW} ^{0.47}} \right),
\end{equation}
where $F_\mathrm{LW}$ is the LW-flux in units of $10^{-21}\mathrm{erg}\ \mathrm{s}^{-1} \mathrm{cm}^{-2} \mathrm{Hz}^{-1}$. The choice for the remaining, crucial parameter of our model $\eta _*=0.01$ will be justified in Section~\ref{sec:tau} by requiring to fit the optical depth to Thomson scattering measured by the \citet{planck13}. The more customary definition of star formation efficiency (SFE), namely the fraction of total gas mass that turns into stars, is related to our efficiency factors by $\eta_\mathrm{eff}=\eta _* f_\mathrm{LW}$.

In our model, we statistically assign a varying number of stars with specific masses to each individual halo, whereas previous studies like \citet{trenti09} or \citet{krhv13} average the quantities like metal yields or amount of ionising photons over the IMF in their model. In order to do so, we randomly select stellar masses from a flat distribution between $M_\mathrm{min}$ and $M_\mathrm{max}$ in each Pop~III forming halo, so that the overall IMF follows Equation~\ref{eq:IMF}. The assignment of stars to a halo is complete, once the total stellar mass exceeds $M_*$. Depending on the dark matter mass and on how far the last star overshoots this criterion, the individual systems contain different amounts of stellar mass, which reflects the stochastic nature of star formation.
The assignment of stars happens instantaneously after virialisation of the halo and we neglect the actual free-fall time of the gas. However, this effect might only delay the whole star formation history in all haloes by about the same time, which is negligibly small on the considered cosmological scales.

\subsection{Pop I/II Star Formation History}
\label{sec:PopIISFH}
Besides an accurate treatment of Pop~III star formation, we also have to model the global star formation history (SFH) and the contribution of Pop I and Pop II stars to the reionisation and metal enrichment of the Universe. Observational constraints on the global SFH are provided by \citet{hb06}, \citet{l08}, and references therein. Based on these observations, \citet{md14} determine the cosmic SFH
\begin{equation}
 \Sigma _{\rm SFH} (z) = \frac{0.015(1+z)^{2.7}}{1+[(1+z)/2.9]^{5.6}} M_\odot \mathrm{\;yr}^{-1} \mathrm{\;Mpc}^{-3},
\end{equation}
where they assume a Salpeter IMF \citep{s55} to convert UV-luminosities into instantaneous star formation rate densities. However, this formula is only valid for $z \lesssim 6$, because there are only few, indirect observational constraints for higher redshift. Hence, for higher redshifts, we use the SFH by \citet{cmsc11}, who modelled the transition of Pop~III to Pop I/II star formation, based on a cosmological simulation.

\subsection{Terminating Primordial Star Formation}
Not all haloes that fulfil the mass criteria (Eq.~\ref{eq:mcrit} and \ref{eq:mdyn}), will form Pop~III stars. Feedback effects like radiation and metal enrichment influence star formation. Whereas radiation has a direct impact on star formation, chemical feedback acts indirectly, by reducing the amount of pristine gas, thus shifting the balance of star-formation modes to the less efficient Pop~II. A more detailed discussion of feedback effects in semi-analytical models of primordial star formation can be found, e.g., in \citet{trenti09,salvadori12,salvadori14}.
We need to know the lifetimes of stars in order to predict when a certain star explodes as a supernova, thus enriching the surrounding medium, and to determine which stars actually survive until today. We interpolate between the lifetimes of non-rotating, metal-free stars provided by \citet{mgcw01} ($0.7-100 M_\odot$), \citet{s02} ($5-500 M_\odot$, without mass loss), and \citet{eetal08} ($9-200 M_\odot$). For masses for which several authors provide a value, we use the mean value. For rotating stars, which we analyse separately later in the paper, we use the values by \citet{eetal08} for the mass range $9-200 M_\odot$. Pop~III stars form predominantly between $z\simeq 15-30$, which corresponds to cosmic ages of $t=100-272$\,Myr. Given the current age of the Universe of $t_{\rm H}=13.8$\,Gyr \citep{planck13}, survivors should have at least lifetimes of $13.54-13.70$\,Gyr. This corresponds to survival masses of $\sim 0.8 M_\odot$.

\subsubsection{Chemical Feedback}
The first stars enrich their surroundings with metals and consequently shut off the formation of subsequent Pop~III stars in these regions. First, we want to focus on the metal yields and the polluted volume by a single star as a function of time. Therefore, we use a simple model for the evolution of the supernova remnant \citep{t50,s59,d11} to determine the time evolution of the SN-enriched volumina. The blast-wave radius as a function of time is given by
\begin{equation}
 R(t) \propto \begin{cases}
 t & \mbox{for } t \leq t_1 \\
 t^{2/5} & \mbox{for } t_1 < t < t_2 \\
 t^{2/7} & \mbox{for } t \geq t_2
 \end{cases}
\end{equation}
with
\begin{equation}
\begin{split}
 t_1 &= 186\mathrm{\;yr} \left( \frac{M_\mathrm{ej}}{M_\odot} \right) ^{5/6} \left( \frac{E_{\rm SN}}{10^{51}\mathrm{erg}} \right) ^{-1/2} \left( \frac{n_0}{0.1 \mathrm{cm}^{-3}} \right) ^{-1/3}\\
%\end{equation}
%\begin{equation}
 t_2 &= 4.93 \times 10^4 \mathrm{\;yr} \left( \frac{E_{\rm SN}}{10^{51}\mathrm{erg}} \right) ^{0.22} \left( \frac{n_0}{0.1 \mathrm{cm}^{-3}} \right) ^{-0.55}
\end{split}
\end{equation}
and
\begin{equation}
\begin{split}
R(t_1) &= 1.9 \mathrm{pc} \left( \frac{M_\mathrm{ej}}{M_\odot} \right) ^{1/3} \left( \frac{n_0}{0.1 \mathrm{cm}^{-3}} \right) ^{-1/3}\\
%\end{equation}
%\begin{equation}
 R(t_2) &= 23.7 \mathrm{pc} \left( \frac{E_{\rm SN}}{10^{51}\mathrm{erg}} \right) ^{0.29} \left( \frac{n_0}{0.1 \mathrm{cm}^{-3}} \right) ^{-0.42}
\end{split}
\end{equation}
where $M_\mathrm{ej}$ is the mass of ejecta, $E_{\rm SN}$ is the explosion energy, and $n_0 \simeq 0.1 \mathrm{cm}^{-3}$ is the number density of the surrounding medium \citep[compare e.g.][]{mesler14}. We assume an explosion energy of $E_{\rm SN}=1.2 \times 10^{51}\mathrm{erg}$ for all SNe, use the ejecta masses by \citet{hw02}, and the individual metal yields by \citet{hw10}. The ongoing expansion for $t \gg t_2$ should mimic the diffuse mixing of metals, which takes over for later times of the expansion.

However, the Universe was polluted simultaneously by many stars and although metal mixing is a highly complex and nonlinear process, we use a simplified statistical model for the enrichment with heavy elements \citep[following][]{kjb08} in order to distinguish whether a star forms in pristine or in previously enriched gas.
Since no metal-free star is allowed to form in the previously enriched vicinity of another star, their spatial distribution is not random, but rather anticlustered. On the other hand, multiple SNe might explode in the same minihalo \citep{ritter14} and the enriched volumina clearly overlap. Assuming that these two effects cancel each other out, we expect a random spatial distribution of SNe and the probability that a specific region has already been affected by $k$ SNe follows the Poisson distribution
\begin{equation}
 P(k,\bar{V}(t)) = \mathrm{e}^{-\bar{V}(t)} \frac{\bar{V}(t) ^k}{k!}.
\end{equation}
The mean value
\begin{equation}
 \bar{V}(t) = \frac{V_\mathrm{enr}(t)}{V_\mathrm{phys}(t)} = \frac{ \sum _i V_\mathrm{SN,i}(t)}{V_\mathrm{phys}(t)},
\end{equation}
represents the dimensionless sum of all metal-enriched volumina divided by the physical volume of the Milky Way. Note that this value can be larger than unity for later times, although there might still be regions with unpolluted gas. The probability for a star to form in pristine gas is therefore given by $P(0,\bar{V}(t))$. If for a random number $0<r<1$, $P(0,\bar{V}(t)) > r$, Pop~III star formation is suppressed for this halo. We also track all haloes in the merger tree that have already experienced metal enrichment by a Pop~III SN, and suppress subsequent primordial star formation in this halo and in all its descendants. Our simple model for metal mixing is consistent with the redshift evolution of the metal volume filling factor by \citet{pallottini14}, who simulate cosmic metal enrichment by the first galaxies. Only for $z \lesssim 7$, our model yields lower values for the fraction of the metal polluted volume, because we do not account for the contribution of Pop I/II star formation 
to this volume. However, in this regime we hardly form any new Pop~III stars anyway.

\subsubsection{Radiative Feedback}
Pop~III stars are more massive and therefore produce more high-energy photons than their present-day counterparts. Hence, they significantly contribute to reionisation of the Universe with ionising photons but also cause photodissociation of H$_2$ with photons in the energy range $11.18-13.6$eV, the so called LW bands. Since the Universe is optically thin to these photons, they build up a background radiation field in the early Universe that influences primordial star formation by removing the most important coolant \citep{har00,mba01,wa07,on08}.
We use the tabulated spectra of \citet{s02}, who provides the production rate of ionising photons, $\dot{N}_\mathrm{ion}$, and the rate of photons in the LW-bands, $\dot{N}_\mathrm{LW}$, for metal free stars between $5-500 M_\odot$. The latter is related to the flux by
\begin{equation}
 F_\mathrm{LW}(z) = c \frac{h \bar{\nu}}{\Delta \nu V_\mathrm{com,halo}} \sum _\mathrm{haloes} \dot{N}_\mathrm{LW,i}(z) t_{*,i},
\end{equation}
where $\Delta \nu = 5.6 \times 10^{14}$Hz is the width of the LW-bands, $h \bar{\nu}=1.98 \times 10^{-11}$erg is the average energy of a LW-photon, $t_{*,i}$ is the lifetime of the $i$-th star, and we sum for each redshift over all contributing Pop~III stars. Here, we implicitly assume that the escape fraction of LW photons from minihaloes is $1.0$, regardless of the minihalo mass or stellar mass. Generally, this escape fraction can be much smaller \citep{kitayama04}, but our conservative assumption tends to reduce the number of Pop~III survivors by radiative feedback and hence strengthens our final conclusions.
Besides LW-feedback, reionisation can also suppress Pop~III star formation in low-mass haloes. However, this effect is only important at smaller redshifts, when hardly any Pop~III stars can form anyway. Since this effect will not change our final conclusions, we do not include this feedback mechanism in the current study.

Pop III star formation likely also leads to the production of a significant soft X-ray      background in the high redshift Universe \citep{oh01,glover03}. The additional ionization produced by these X-rays catalyzes H$_{2}$ formation, and in the absence of a LW background can exert a positive feedback on Pop III star formation \citep{haiman96b}. However, when both LW photons and X-rays are present, negative feedback from the LW photons generally dominates over the positive feedback from the X-rays \citep{glover03,machacek03}, and so accounting for the effects of the X-ray background would not significantly change the results of our model.

\section{Empirical Constraints}
In this section, we justify our choice of parameters by comparing several model predictions to existing observations. All these empirical constraints assume that a Milky Way-like halo is representative on cosmological scales and hence that high-redshift observations reflect the state of Milky Way progenitors at earlier times. Moreover, we use these constraints to calibrate the Pop~III SFE.

\subsection{Optical Depth to Thomson Scattering}
\label{sec:tau}
The optical depth to Thomson scattering is the most important constraint, because it can be determined fairly accurately and, together with its errorbars, it yields an upper and lower limit to the star formation rate, whereas the other two empirical constraints only yield upper limits. Following \citet{retal13}, the optical depth is given by
\begin{equation}
 \tau = c \sigma _T n_H \int _0 ^z \mathrm{d}z' f_e Q_{\rm ion} (z') (1+z')^3 \left| \frac{\mathrm{d}t}{\mathrm{d}z'} \right|,
\end{equation}
where $z$ is the redshift of emission, $\sigma _T = 0.665 \times 10^{-24} \mathrm{cm}^2$ the cross-section to Thomson scattering, $n_\mathrm{H}$ the comoving hydrogen number density, $Q_{\rm ion}$ the volume filling fraction of ionised regions, and $f_e$ the number of free electrons per hydrogen nucleus (singly/doubly ionised helium) in the ionised intergalactic medium (IGM)
\begin{equation}
 f_e = \begin{cases}
 1+Y_p/2X_p & \mathrm{at } z \leq 4 \\
 1+Y_p/4X_p & \mathrm{at } z > 4,
 \end{cases}
\end{equation}
where $Y_p$ and $X_p$ are the primordial abundances of He and H, respectively. The time evolution of the volume filling fraction is based on
\begin{equation}
 \frac{\mathrm{d} Q_{\rm ion}(z)}{\mathrm{d}z} = \frac{1}{n_H} \frac{\mathrm{d} n_{\rm ion}(z)}{\mathrm{d}z} - \frac{Q_{\rm ion}(z)}{t_{\rm rec}(z)} \left| \frac{\mathrm{d}t}{\mathrm{d}z} \right|
 %\alpha _B n_b C(z) Q_{ion} ^2 (z) (1+z)^3 \left| \frac{\mathrm{d}t}{\mathrm{d}z} \right|
  \label{eq:Qion2}
\end{equation}
Here, $n_{\rm ion}$ is the comoving number density of ionising photons, and the average recombination time in the IGM 
\begin{equation}
 t_{\rm rec} = [ C_\mathrm{HII} \alpha _B (1+Y_p/4X_p) n_H (1+z)^3 ]^{-1},
\end{equation}
where $\alpha _B = 2.6 \times 10^{-13} \mathrm{cm}^3 \mathrm{s}^{-1}$ is the case B recombination coefficient, and $C_\mathrm{HII} = 3$ is the clumping factor \citep{retal13}. For the number density of ionising photons we have to distinguish the Pop~III and Pop I/II cases:

\subsubsection{Pop I/II}
We model the SFH $\Sigma _\mathrm{SFH} (z)$ as described in Section~\ref{sec:PopIISFH}, and determine the comoving number density of ionising photons by
\begin{equation}
 \frac{\mathrm{d} n_{\rm ion}(z)}{\mathrm{d}z} = f_{\rm esc} \eta _{\rm ion} \frac{\Sigma_{\rm SFH} (z)}{\mu m_{\rm H}} \left| \frac{\mathrm{d}t}{\mathrm{d}z} \right| ,
 \label{eq:nion2}
\end{equation}
where $f_{\rm esc}=0.3$ is the escape fraction of ionising photons \citep{gb06,retal13}, and $\eta _{\rm ion}=4.0 \times 10^3$ the number of ionising photons emitted per stellar baryon \citep{gb06}. The escape fraction for the present-day Galaxy is much smaller ($\sim 6\%$, \citet{bland99}) than it was at earlier times. This is mainly due to the fact the the stellar feedback is more efficient in lower mass halos in clearing away the gas. Since there is no unique escape fraction that is constant in space and time \citep{pkd13}, we use this single value as an average for our simplified model. The resulting optical depth for the Pop~I/II-only model is
\begin{equation}
 \tau _{\rm PopI/II} = 0.065 .
\end{equation}
This is the baseline contribution from known stellar populations, and it is
evident that hitherto undetected sources at high redshifts are needed
to provide the balance.

\subsubsection{Pop~III}
For the merger tree, we can directly assign the number of ionising photons to each Pop~III star based on the values by \citet{s02}. The time-averaged escape fraction of ionising photons in Pop~III-forming haloes is $f_{\rm esc,III}=0.7$ \citep{gb06}. This Pop~III contribution is added to the baseline contribution by Pop~I/II stars (see Eq.~32).

\subsubsection{Calibrating Pop~III Star Formation Efficiency}
\begin{figure}
\centering
\includegraphics[angle=-90]{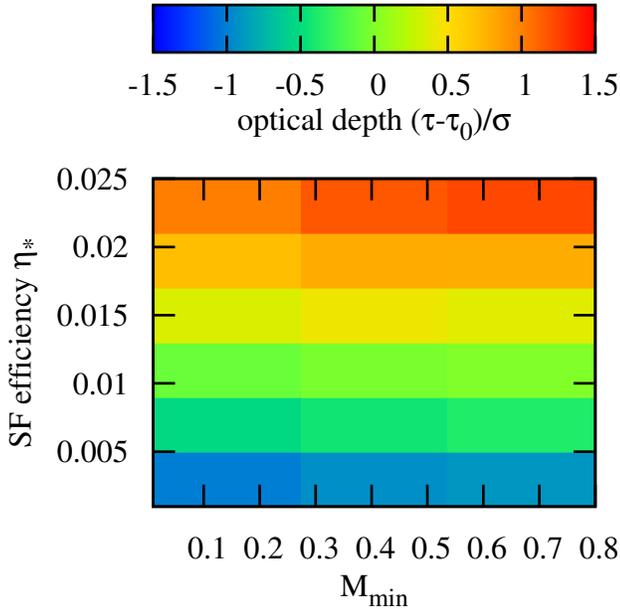}
\caption{Optical depth minus the \citet{planck13} value normalised to the $1\sigma$ error as a function of the SFE and of the lower IMF mass limit. Since ionising photons are mainly produced by high-mass stars, the lower IMF mass limit hardly affects the optical depth. However, the optical depth crucially depends on the SFE and is therefore used to calibrate this key model parameter.}
\label{fig:etaMmintau}
\end{figure}

\begin{figure}
\centering
\includegraphics[angle=-90]{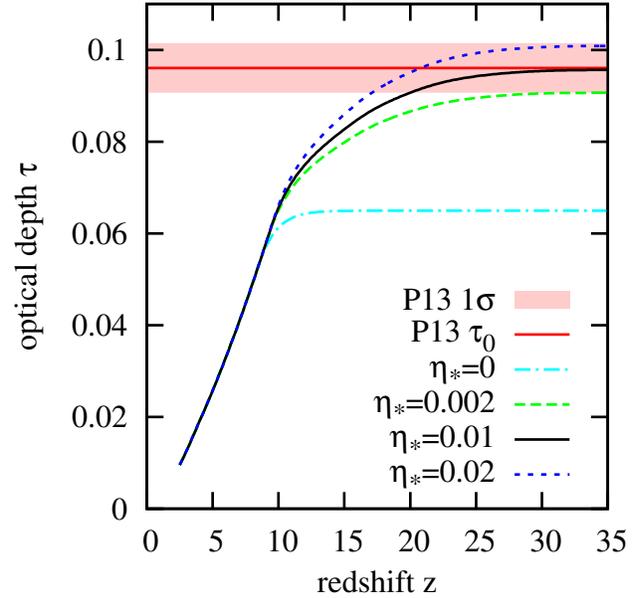}
\caption{Optical depth as a function of redshift for different choices of the SFE $\eta _*$ and comparison to the observational constraint by the \citet{planck13}. The contribution from Pop~I/II star formation yields a baseline value of $\tau _{\rm PopI/II} = 0.065$, and the complete model results in values in accordance with the observational constraint for the three selected star formation efficiencies.}
\label{fig:tau}
\end{figure}

The \citet{planck13} value for the Thomson optical depth with its $1\sigma$ error is $\tau_0 = 0.0961 \pm 0.0054$, which yields a possible range of
\begin{equation}
 0.0907 \leq \tau \leq 0.1015.
\end{equation}
This value constrains the number of ionising photons produced in our model, which is critically sensitive to the SFE, $\eta _*$, of Pop~III stars. Consequently, we will use three different values of this efficiency for our further studies, $\eta _\mathrm{min}$, $\eta _\mathrm{best}$, and $\eta _\mathrm{max}$, which reproduce the lowest possible, the best, and the highest possible estimate of the optical depth. The general dependence of the optical depth on the SFE and on the lower IMF mass limit can be seen in Figure \ref{fig:etaMmintau}.
The optical depth depends critically on the SFE, but hardly on the choice of $M_{\rm min}$. Consequently, we can fix the SFE for our fiducial model ($M_{\rm min} =0.01 M_\odot$), which yields values for $\eta _*$ in the range of $\eta _\mathrm{min}=0.002$ to $\eta _\mathrm{max}=0.02$ with a best fitting value of $\eta _\mathrm{best}= 0.01$. The corresponding optical depths as a function of redshift are displayed in Figure \ref{fig:tau}. We emphasise that this Pop~III SFE calibration is not very sensitive to the chosen value of the lower mass limit, which allows us to vary this lower limit during our analysis, without altering the SFE.

\subsection{Metal Enrichment}
\label{sec:zZ}
\begin{figure}
\centering
\includegraphics[angle=-90]{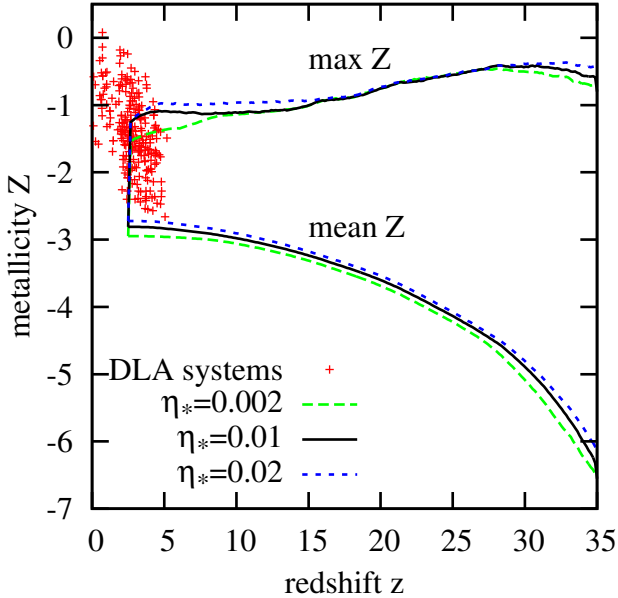}
\caption{Mean and maximal metallicities in the proto Milky Way-like halo, as a function of redshift, for different star formation efficiencies. The model predictions are compared to observed DLA metallicities \citep{retal12}. Evidently, we can conclude that there were already sufficiently enriched regions at high redshifts to explain the DLA metallicities, which in turn supports our metal enrichment model.}
\label{fig:zZ}
\end{figure}

Observations of damped Ly$\alpha$ systems (DLAs) provide gas-phase metallicities at large cosmological look-back times with high precision. Hence, we use the chemical enrichment of DLAs as a constraint on the metal enrichment history of the Milky Way, under the assumption that our Galaxy went through a DLA-like phase earlier on \citep{krhv13,khrv14}. \citet{retal12} provide a compilation of DLA metallicities up to redshift $z\simeq 5$, which our model should be able to reproduce. The metal enrichment is very inhomogeneous and the question of whether one can find an overdensity with a certain metallicity at a specific redshift is fundamentally a statistical one. Hence we plot, for any redshift, the mean and the maximum metallicity, which should bracket the observed metallicities. After virialisation of the Milky Way ($z \simeq 2.5$), we only have one halo and consequently, mean and maximum metallicity are the same. This is related to our very simple estimation of the metallicity, which is 
just based on the mass of the halo and the mass of contained metals. However, we should keep in mind that this final target halo in our merger tree actually consists of 
many subhaloes, which already had higher individual metallicities at earlier times. The expectation, therefore, is that
there must be haloes at sufficiently high redshifts that have exceeded the DLA metallicities, and finally end in the Milky Way. The metal enrichment history is illustrated in Figure \ref{fig:zZ}, which shows that this requirement is fulfilled.

\subsection{Black Holes and Unresolved X-Ray Background}
\label{sec:uxrb}
\begin{figure}
\centering
\includegraphics[angle=-90]{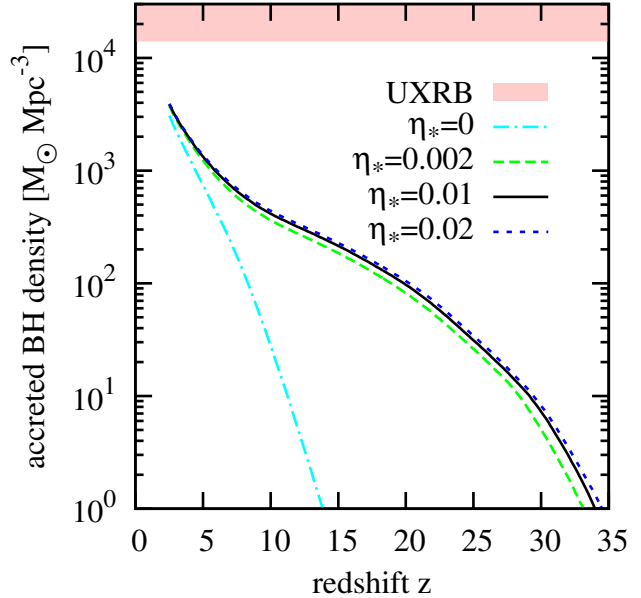}
\caption{Mass density accreted onto black holes as a function of redshift and upper limit based on the observationally inferred UXRB. It is evident that our model does not violate this empirical constraint.}
\label{fig:UXRB}
\end{figure}

The formation and accretion histories of supermassive black holes are still not completely understood \citep{v12}, and can therefore not be used to test our model. However, measurements of the cosmic X-ray background can be used to constrain the population of high-redshift X-ray sources \citep{mirabel11}. The strength of the unresolved X-ray background (UXRB), yields an upper limit for the mass accreted by BHs above $z \geq 5$, which should remain below $1.4 \times 10^4 M_\odot \mathrm{Mpc}^{-3}$ for $z>5$ \citep{metal12,shvm12}. In order to predict the corresponding contribution to the UXRB, we follow \citet{jpbm14} and assume that $30\%$ of BHs evolve into a high-mass X-ray binary \citep{pwcw09}, which accretes gas from the stellar companion at the Eddington rate of $2.2 \times 10^{-6} M_\odot \mathrm{yr}^{-1} (M_{\rm BH}/100 M_\odot)$ for a duration of 2\,Myr each. Afterwards, BHs accrete diffuse halo gas with the Bondi-Hoyle accretion rate, which varies between $10^{-14} - 10^{-6} M_\odot \mathrm{yr}
^{-1}$, depending on the conditions near the BH. According to 
\citet{awa09}, 
this value is between $10^{-12}-10^{-9} M_\odot \mathrm{yr}^{-1}$ with a mean of about $10^{-10} M_\odot \mathrm{yr}^{-1}$. This latter value also appears to be an upper limit for the accretion rates in \citet{jpbm14}. The accreted mass per Pop~III-forming minihalo is therefore given by
\begin{equation}
 M_{\rm acc,BH}(z) = \int _{z_\mathrm{form}} ^{z} 10^{-10} M_\odot \mathrm{yr}^{-1} \frac{\mathrm{d}t}{\mathrm{d}z} dz,
\end{equation}
where $z_\mathrm{form}$ is the formation redshift of the black hole. We estimate the baseline Pop~I/II contribution from the corresponding SFH, together with the assumption that $0.8\%$ of the mass ends up in stellar-mass BHs, according to a Kroupa IMF \citep{k01}. However, one should keep in mind that the UXRB constraint could only falsify our model in case that it were to overproduce X-rays, thus violating the empirical UXRB upper limit. As is evident from Figure~\ref{fig:UXRB}, our model passes this consistency check.

\subsection{Hints from Metal Poor Stars}
Our standard model assumes a logarithmically flat IMF from $M_\mathrm{min}=0.01 M_\odot$ to $M_\mathrm{max}=100 M_\odot$. However, in the following sections we will vary these limits and test the sensitivity of the number of Pop~III survivors with respect to these parameters.
For the lower IMF limit, we explore the range between the opacity limit of $0.01 M_\odot$ and the survivability threshold of $0.8 M_\odot$. If the true $M_{\rm min}$ were in excess of $0.8 M_{\odot}$, there would evidently exist no Pop~III survivors in the local Universe.
For the high-mass end of our IMF, we have to be able to create primordial stars with at least $60 M_\odot$, the Pop~III progenitor mass implicated in producing the metals locked up in the most iron-poor star discovered so far \citep{ketal14}. On the other hand, following \citet{kjb08}, we should limit the upper end of the IMF to $170 M_\odot$, in order to not have more than $7\%$ of Pop~III stars that end as PISNe \citep[see also][]{aoki14}.

\section{Model Predictions}
In this section, we present the main results of our analysis. We first discuss the Pop~III star formation history and related quantities, such as the build-up of LW radiation, and the different mechanisms that act to suppress Pop~III star formation. Subsequently, we investigate the stellar-archaeological constraints on the primordial IMF, and specifically assess the observational sample sizes required to constrain its lower mass limit.

\subsection{History of Pop~III Star Formation}
\begin{figure}
\centering
\includegraphics[angle=-90]{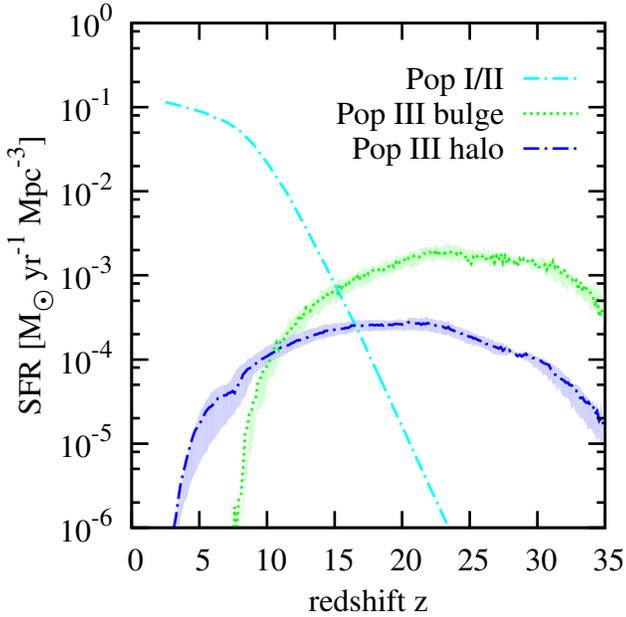}
\caption{Star formation rate as a function of redshift for Pop~III stars in the bulge and in the halo as a function of redshift. The lines represent the standard model ($\eta _* = 0.01$), whereas the shaded areas indicate the possible variations by using $\eta _* = 0.002$ and $\eta _* = 0.02$. The Pop~I/II SFH is also plotted for comparison. Notice that Pop~III star formation in the bulge peaks earlier and is about an order of magnitude higher than in the halo before $z \simeq 20$. Due to the much smaller volume of the bulge, the overall rate of Pop~III star formation is dominated by the halo contribution.}
\label{fig:SFR}
\end{figure}

Whereas the Pop I/II star formation history is based on analytical formulas, we model Pop~III star formation self-consistently with the most relevant feedback mechanisms taken into account. Moreover, our distinction between the Milky Way bulge and halo allows us to point observers to the most promising region. For each set of model parameters, we create 25 different merger tree realisations and consequently 25 slightly different merger histories of the Milky Way. The final parameters, which are presented in this section, are averaged over these merger tree realisations. The star formation rates for our fiducial model are shown in Fig.\;\ref{fig:SFR}.
The star formation rate in the bulge peaks earlier and is higher at earlier times, which is in agreement with the inside-out growth of galaxies. Following our treatment, the bulge is the first massive object that formed and therefore consists of many haloes that have virialised very early in time. Consequently, Pop~III star formation in the bulge is not as influenced by suppressing feedback mechanisms as the halo. However, the comoving volume of the bulge is very small compared to the halo and therefore, the overall Pop~III star formation rate is dominated by the halo contribution.

\begin{figure}
\centering
\includegraphics[angle=-90]{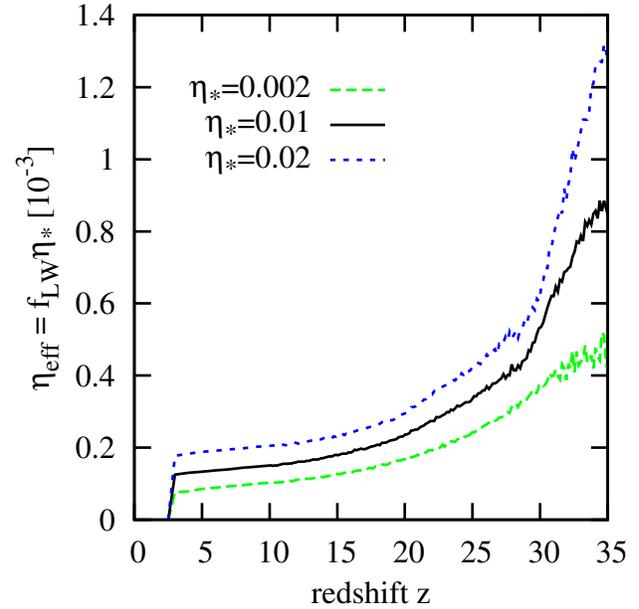}
\caption{Overall Pop~III SFE as a function of redshift. The parameter $\eta _\mathrm{eff}$ combines the two individual efficiencies in Equation~\ref{eq:etas}, and indicates which fraction of the total baryonic mass in the halo, on average, turns into Pop~III stars.}
\label{fig:eta_eff}
\end{figure}

\begin{figure}
\centering
\includegraphics[angle=-90]{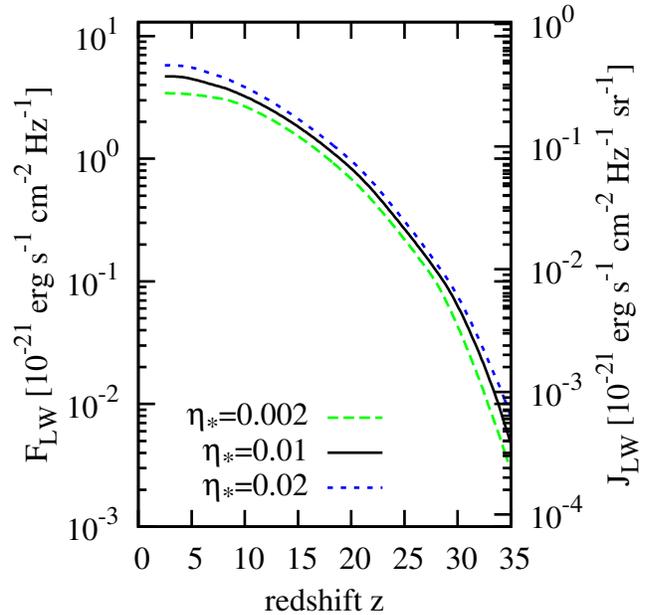}
\caption{Average background flux of LW-photons as a function of redshift. Our model prediction is consistent with the results by \citet{jgb08}.}
\label{fig:FLW}
\end{figure}

Evidently, a key parameter is the overall, halo-scale Pop~III SFE, i.e. the fraction of baryonic matter that turns into Pop~III stars in each dark matter host halo. This parameter is determined by the ability of the gas to cool, and therefore depends on its chemical composition and the presence of external radiation backgrounds. The resulting effective (overall) star formation efficiencies in our model are shown in Fig.\;\ref{fig:eta_eff}.
The decrease in SFE from $10^{-3}$ to $10^{-4}$ between $z \simeq 35$ and $z \simeq 10$ is mainly driven by the increasing LW background (see Fig.\;\ref{fig:FLW}), such that the associated photodissociation of H$_2$ limits the ability of the primordial gas to cool.

\begin{figure}
\centering
\includegraphics[angle=-90]{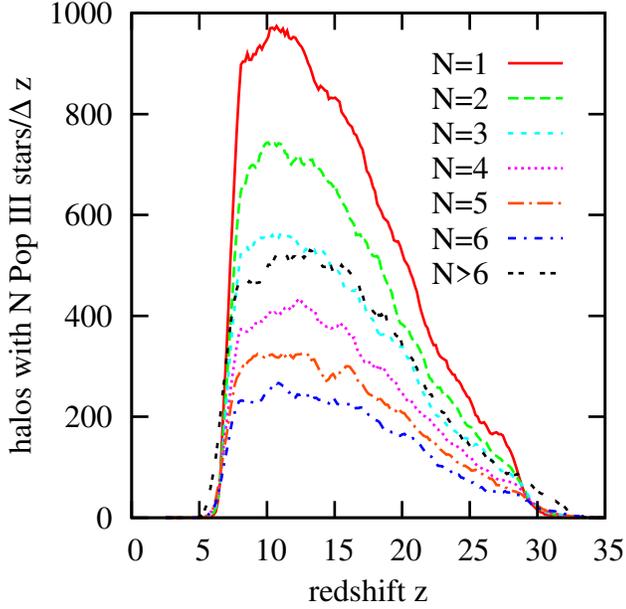}
\caption{Number of Pop~III stars per minihalo for our standard model. At any redshift, this plot illustrates how many new haloes form with $N$ Pop~III stars per unit redshift bin. Although we use a simple probabilistic IMF sampling, this approach reproduces the results of numerical simulations that most of the primordial stars form in higher multiple systems.}
\label{fig:multipli}
\end{figure}

\begin{figure}
\centering
\includegraphics[angle=-90]{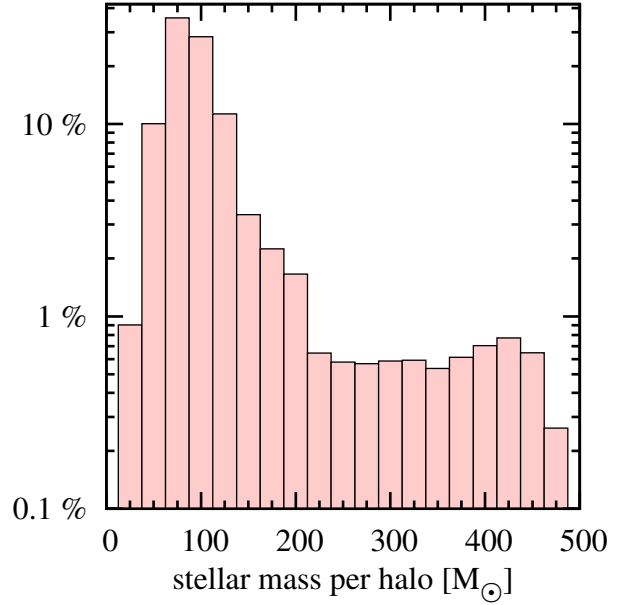}
\caption{Distribution of the stellar mass per halo. Most of the haloes have a stellar mass content of $\sim 100 M_\odot$ and the broad distribution from $\sim 25 M_\odot$ to $\sim 500 M_\odot$ reflects the stochastic nature of Pop~III star formation and disc fragmentation.}
\label{fig:IMFhisto}
\end{figure}

A related question concerns the fragmentation and the number of primordial stars per halo. We assign individual stars to each Pop~III-forming halo, until we reach the desired total stellar mass in this halo. The corresponding number of stars per halo as a function of redshift can be seen in Fig.\;\ref{fig:multipli}.
In our model, only some systems contain one Pop~III star, whereas the majority of the systems contains binary or higher multiple systems. Some rare systems host even up to 19 Pop~III stars. These multiplicities are not based on a detailed three-dimensional disc fragmentation simulation, but are rather based on a probabilistic assignment. Phrased differently, we do not mimic mergers or ejections of primordial stars, but assign the final number of stars to the system. Although the outcome agrees with simulations, this distribution of multiplicities is not physically motivated. The associated stellar mass per halo can be seen in Fig.\;\ref{fig:IMFhisto}.
Following our recipe of Pop~III star formation, we have a broad range of stellar masses per halo from $\sim 25 M_\odot$ to $\sim 500 M_\odot$ with a mean around $\sim 100M_\odot$. This distribution complies with the expectation from numerical simulations of disc fragmentation in primordial gas clouds.

\begin{figure}
\centering
\includegraphics[angle=-90]{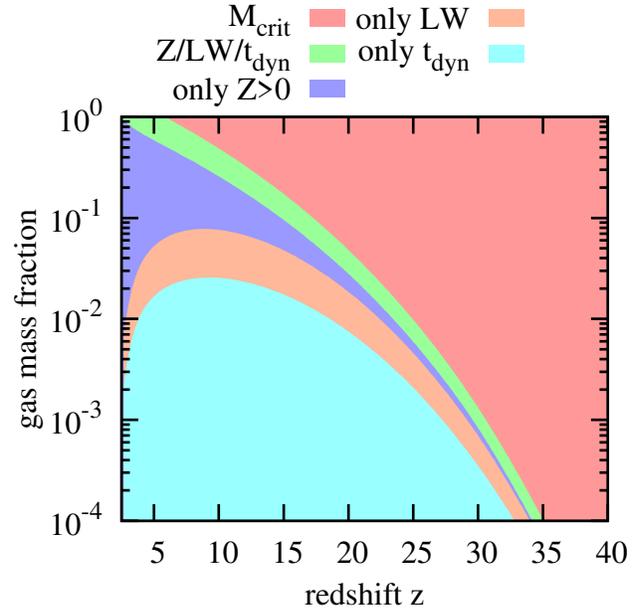}
\caption{Relevance of different feedback and suppression mechanisms for Pop~III star formation. At any redshift, this plot illustrates the dominant suppression effects, quantified by their respective gas mass fractions. $M_\mathrm{crit}$ reflects gas that is not part of sufficiently massive haloes. The green area indicates the gas that does not form Pop~III stars because of suppression by several mechanisms, whereas the purple, magenta, and blue areas represent gas that cannot form Pop~III stars only because of previous metal enrichment, too strong a LW-flux, or dynamical heating, respectively.
Keeping in mind the logarithmical scaling, it can be seen that the only relevant suppression mechanisms are metal enrichment at smaller redshifts and the critical mass criterion earlier on.}
\label{fig:supp}
\end{figure}

Regarding the termination of Pop~III star formation, we are interested in the different feedback mechanisms and their individual importance. In our model, dark matter haloes have to fulfil four criteria in order to form Pop~III stars: they have to be massive enough, should not be affected by dynamical heating due to mergers, should not have been polluted by metals, and should not experience too strong a LW background.
In our merger-tree algorithm, we do not resolve haloes below $6 \times 10^4 M_\odot$, and therefore cannot assess the detailed physics of suppressing star formation inside these haloes. However, for all other haloes, we can explicitly identify whether one effect dominates, or whether several criteria suppress star formation simultaneously. The time-dependent relevance of different suppression mechanisms is shown in Fig.\;\ref{fig:supp}.
The critical mass is obviously the dominant suppression mechanism most of the time. Only at smaller redshifts, other feedback mechanisms become important. Whereas there are hardly any haloes in which Pop~III star formation is suppressed only because of the LW-background or dynamical heating, there is quite a number of haloes that cannot form Pop~III stars because of metal enrichment or a combination of these feedback mechanisms.

\subsection{Stellar Archaeology}
We can constrain the lower mass limit of the primordial IMF from the fact that we have not observed any Pop~III star so far. Moreover, we can wield this null-result into an accurate probe of $M_{\rm min}$ by considering the current survey sizes of extremely metal-poor (EMP) stars. The basic idea here is that the probability for the detection of a Pop~III survivor increases with decreasing $M_{\rm min}$. The ongoing efforts by stellar archaeologists thus provide ever improving empirical upper limits on this survival probability ($P_{\rm surv}<1/N_{\rm sample}$). The exciting prospect then arises that even the non-detection of a Pop~III survivor can constrain the primordial IMF at a level of precision that is otherwise completely out of range. In the following, we discuss this IMF probe in detail.

\subsubsection{Current Sample Size}
Our statistical analysis is based on a combinatorial argument. Hence, we need to know the current sample size, $N_o$, of randomly chosen stars, from which candidates have been photometrically selected for follow-up spectroscopic measurement of their metallicity.
To detect a primordial star, it has to show no sign of metal lines of any kind in its stellar spectrum. The most prominent lines used to determine metallicity are the Ca{\sc ii}, K, or Fe{\sc i} lines \citep{caffau11,caffau13}. The Fe{\sc i} line by itself is not a good indicator, as we know of at least one star which appears to be entirely iron-deficient, but which despite this has a carbon abundance of [C/H]~$\sim -2.5$, and which is therefore not a Pop~III star \citep{ketal14}.
To exclude the primordial composition, we need high-resolution spectra with spectral resolution $R>20,000$ and a signal-to-noise of $S/N>50$, which are currently available for $\sim 1,000$ halo stars and which will serve as a conservative lower value.
However, the Hamburg/ESO survey has observed $\sim 4 \times 10^6$ individual sources (N. Christlieb 2014, priv. comm.), photometrically selected metal-poor candidates and 
spectroscopically measured the actual metallicity of a smaller subset. For our further analysis, we are interested in the number of randomly selected halo stars. Since the original sample of point sources contained also quasars, over-saturated stars or disc stars, the actual number is somewhat lower and we 
use $N_o=10^6$ as a rough upper limit:
\begin{equation}
 10^3 \lesssim N_{o,h} \lesssim 10^6.
\end{equation}
However, one should keep in mind that the selection criteria for follow-up spectroscopy might have rejected a metal-free star \citep{csfbw08}.

For the Milky Way bulge, the number of observed stars is much smaller. There are observations of $100$ EMP stars \citep{gpetal2013}, which were selected from a slightly bigger sample and will serve as a lower estimate. The ARGOS survey, however, obtained spectra for $N_o = 28,000$ stars at a spectral resolution of $R=11,000$, which will serve as an upper limit for the bulge \citep{ness13}. Consequently, the current sample size for the bulge is in the range
\begin{equation}
 10^2 \lesssim N_{o,b} \lesssim 2.8\times10^4.
\end{equation}
The actual numbers of the current sample sizes for bulge and halo do not directly enter our statistical model, but should give a rough idea of the effectiveness of continuing observations in these regions.

\subsubsection{Statistical Description}
Since the statistical method is the same for bulge and halo, we will derive it for an arbitrary set of $N_t$ stars in total with $N_o$ observed stars and $N_s$ expected survivors. Note, that $N_o$ is treated as a free parameter, reflecting the increasing sample sizes of upcoming surveys. The total number of different realisations, where we observe $N_o$ out of $N_t$ stars is given by
\begin{equation}
 N_\mathrm{tot} = \binom{N_t}{N_o}.
\end{equation}
The number of different realisations, where we observe $N_o$ out of $N_t$ stars, but do not observe any Pop~III survivor is given by
\begin{equation}
 N_\mathrm{not} = \binom{N_t-N_s}{N_o}.
\end{equation}
The probability that we have missed all expected Pop~III survivors in the observed sample is therefore given by
\begin{equation}
 p_0 = \frac{N_\mathrm{not}}{N_\mathrm{tot}} = \frac{(N_t - N_s)! (N_t-N_o)!}{N_t! (N_t-N_s-N_o)!}.
\end{equation}
In the Milky Way, the values of $N_t$, $N_s$, and $N_o$ are too high to calculate the factorial. Hence, we use Stirling's formula to simplify matters:
\begin{align}
 \ln p_0 = &(N_t-N_s) \log (N_t-N_s) + (N_t-N_o) \log (N_t-N_o) \nonumber \\
           &- N_t \log (N_t) - (N_t-N_o-N_s) \log (N_t-N_o-N_s)
\label{eq:lnp0}
\end{align}
Phrased differently, $p=1-p_0$ describes the probability that the current sample size is representative and we have no Pop~III survivor in the Milky Way.

Based on the lower mass limit for the primordial IMF $M_\mathrm{min}$, our semi-analytic model predicts the number of Pop~III survivors. Using Eq.~(\ref{eq:lnp0}), we can now determine the critical sample size to exclude Pop~III stars with masses below $M_\mathrm{min}$ for three different reliability thresholds. This statistical prescription is valid, as long as the sample is a random, unbiased selection of stars and all stars have the same probability of being observed. For previous surveys, the latter assumption generally breaks down for stellar masses below $\sim 0.6 M_\odot$, because these stars are generally to faint for direct observation (Anna Frebel, priv. comm.). However, this threshold is also a moving target, which might decrease with upcoming surveys.

\subsubsection{Prediction of lower IMF limit}
\begin{figure*}
\centering
\includegraphics[angle=-90]{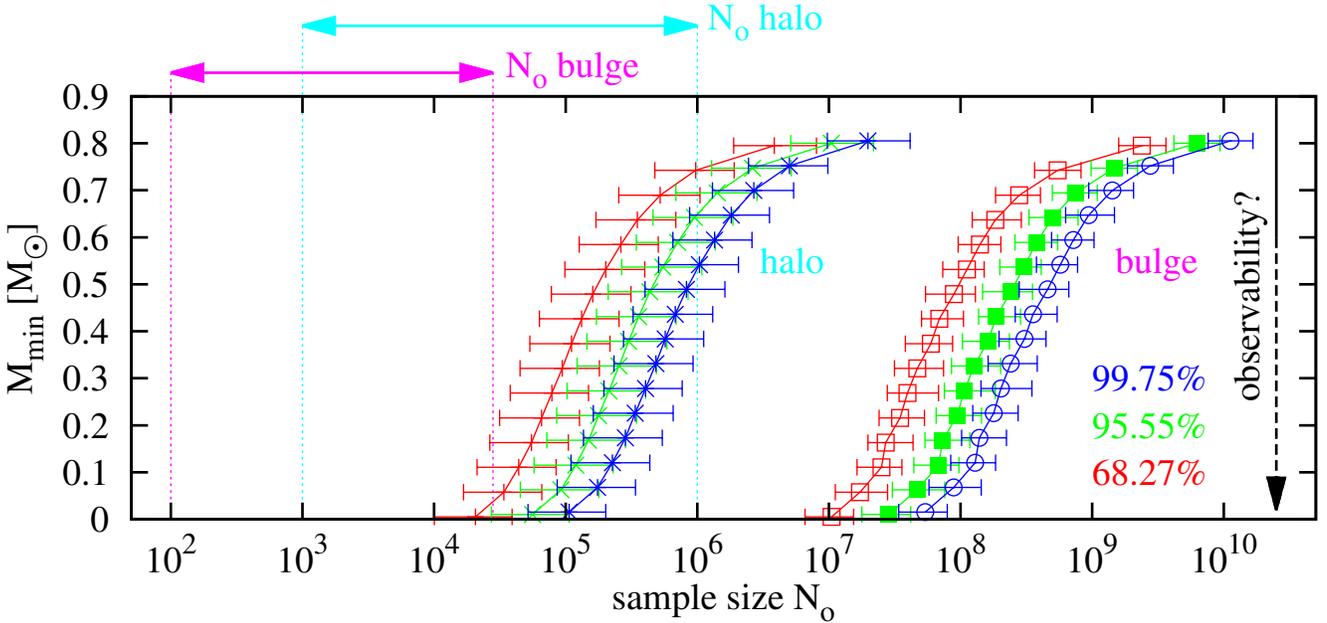}
\caption{Constraints on the lower IMF limit as a function of the sample size, whereas the different lines represent the confidence levels of $68.27\%$, $95.55\%$, and $99.73\%$. The vertical dashed lines in magenta and light blue indicate the current sample sizes, respectively the number of stars that have already been observed and that are certainly not Pop~III stars. The vertical arrow on the right illustrates the reduced observation probability for stars below $\sim 0.6 M_\odot$. However, since we model star formation separately for every value of $M_\mathrm{min}$, values above the observability threshold are not affected by the reduced observability for smaller stellar masses. For the bulge, we have to observe a much higher number of stars to find a constraint on $M_\mathrm{min}$, compared to the halo. Furthermore it is noticeable that we are already in the interesting regime for the halo and we need sample sizes of $4 \times 10^6$, $1 \times 10^7$, $2 \times 10^7$ to exclude any 
Pop~III survivors with a confidence level of $68.27\%$, $95.55\%$, $99.73\%$, respectively. The error bars include the uncertainty in the SFE, in the total number of stars $N_t$ and the statistical scatter between several merger tree 
realisations.}
\label{fig:masterplot}
\end{figure*}

We now proceed to the core of our argument, deriving the sample sizes required
to effectively constrain the lower-mass limit of Pop~III.
An important parameter for our analysis is the total number of stars in the Milky Way bulge and halo. For a characteristic mass of a star of $0.38 M_\odot$ \citep{k01}, total stellar masses of about $10^9 M_\odot$ for the metal-poor halo and $2\times10^{10}M_\odot$ for the bulge we expect $N_{t,h} \simeq 2.5  \times 10^9$ stars in the halo and $N_{t,b} \simeq 5.5 \times 10^{10}$ stars in the bulge. Although these values are crucial for our conclusion, we can only provide an order of magnitude estimate. To illustrate this uncertainty, we assume that both numbers are subject to an error of $25\%$. The resulting constraints on the lower IMF limit as a function of the sample size can be seen in Fig.\;\ref{fig:masterplot}.
There are several important conclusions that we can draw from this plot. The sample size needed for constraining $M_\mathrm{min}$ is more than two orders of magnitude higher for the Milky Way bulge than for the halo. Consequently, observations in the halo are much more promising for constraining the lower mass IMF limit, especially because the current sample size is already higher for the halo and because observations in the bulge are hindered by dust extinction. For an optimistic reading of the sample size of the Hamburg/ESO survey and a conservative treatment of the errorbars, we could already exclude the existence of any Pop~III stars with less than $\sim 0.65 M_\odot$ with a certainty of $95\%$. However, for a more restrictive reading (corresponding to $\sim 10^3$ halo stars surveyed at sufficiently high quality), no constraints could yet be placed on the Pop~III IMF. In order to exclude any Pop~III survivors with a certainty of $99\%$, a critical sample size of $\sim 2 \times 10^7$ halo stars has to be 
achieved, which should be well within reach of upcoming stellar archaeological campaigns. However, designing a well considered observing plan is out of the scope of this work. A basic assumption of this statistical analysis is that all stars have the same probability of being observed. Once this assumption breaks down, we might have to correct for the reduced observation probability. However, any conclusions drawn from the mass range above $\sim 0.65 M_\odot$ is not affected by this caveat.

\section{Caveats and Parameter Sensitivity}
\begin{figure*}
\centering
\includegraphics[angle=-90]{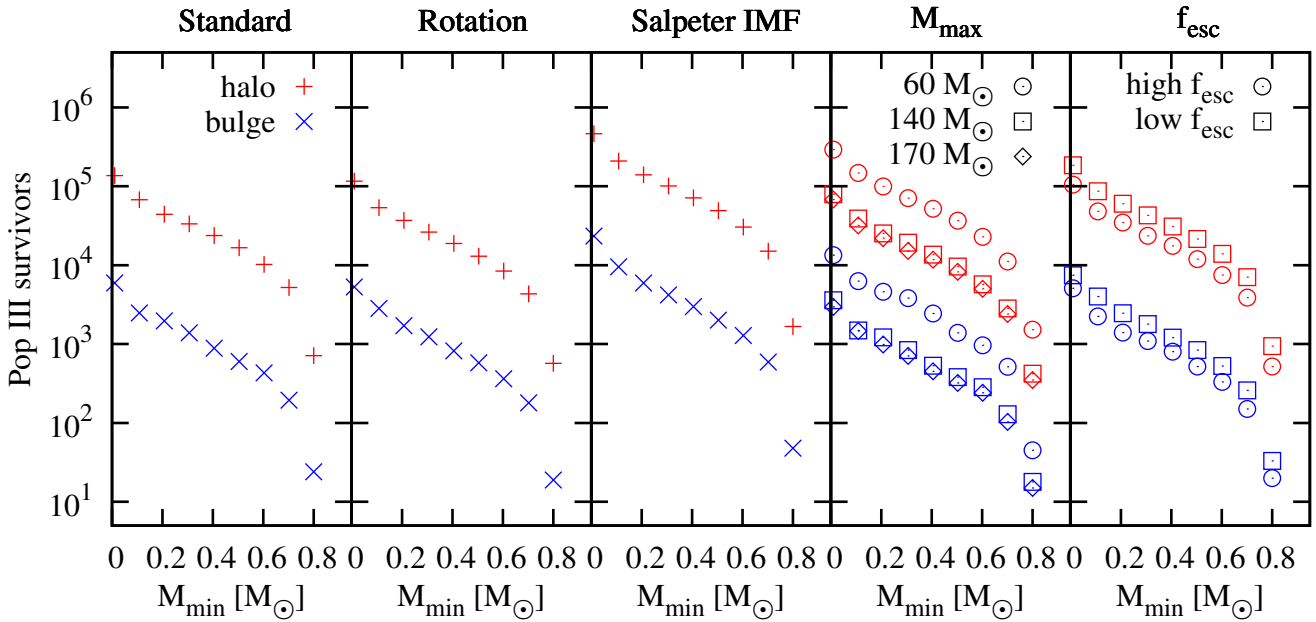}
\caption{Expected number of Pop~III survivors as a function of the lower IMF limit for different assumptions. The standard model assumes non-rotating Pop~III stars with a logarithmically flat IMF, masses up to $100 M_\odot$, and escape fractions of $f_{\rm esc,III}=0.7$ and $f_{\rm esc}=0.3$. The assumption of rotating Pop~III stars increases the stellar lifetimes, hence the production of ionising photons. Consequently, the SFE and the number of survivors is slightly smaller. For a Salpeter IMF, the SFE has to be considerable higher and we expect about an order of magnitude more survivors. The upper mass limit of the primordial IMF is a crucial parameter and, especially for values $<100 M_\odot$, it has a significant influence on the number of survivors, because it determines the gas mass that is left over for possible survivors.
The escape fraction for ionising photons is very uncertain, but even a large change in this parameter hardly influences the number of Pop~III survivors. This parameter study shows that most uncertainties might yield even more Pop~III survivors, which in turn strengthens our final conclusions.}
\label{fig:parameters}
\end{figure*}

Although we include the most relevant feedback mechanisms and calibrate our model against empirical constraints, there are several approximations and limitations that introduce uncertainties to our results. In this section, we investigate these caveats and address the question of how sensitive our model is to the specific choice of parameters. We begin by discussing some of the processes that are not included in our current approach in Section \ref{sec:NeglEff}. It is fair to say that the numerical calculations of Pop~III star formation that aim at resolving individual objects are still in their infancy and leave room for large uncertainties with respect to stellar multiplicity and rotation as well as to the IMF. We assess the influence of these uncertainties in Sections \ref{sec:rotation} and \ref{sec:IMFdep}, respectively. Another key factor that enters our model is the cosmic reionisation history, as modelled by the escape fraction of ionising radiation. The uncertainties in this parameter are explored in 
Section \ref{sec:ReionHist}. We focus our discussion on 
the number of expected Pop~III survivors as our primary prediction, and provide the corresponding plots in Fig.\;\ref{fig:parameters}.

\subsection{Neglected Effects}
\label{sec:NeglEff}
There are a number of physical processes and effects that are not yet included in our model. The first simplification is that our approach is a purely statistical one. We have no information about the exact spatial location of dark matter haloes and the distance to their neighbours. For this reason, we can address the question of how feedback influences other haloes only in a probabilistic and rather idealised fashion. By the same token, we can determine whether a high-redshift halo will become part of the Milky Way halo or the bulge in a statistical sense only. 

Furthermore, we also neglect the effects of streaming velocities \citep{tseliakhovich10} on the dynamics of primordial halos and on their ability to form stars. This effect may lead to a delayed onset of Pop III star formation, possibly occurring in haloes of somewhat larger masses than we assume here \citep[e.g.][]{getal11a, maio11, stacy11}. However, we note that in the model of \citet{tseliakhovich11}, the critical mass for collapse in a typical region is increased by only one order of magnitude and hence remains smaller than our $M_{\rm crit}$ at all redshifts $z < 40$ \citep{g13}. It is therefore plausible that properly accounting for the effects of streaming velocities would not make a major difference to the results of our model.
We also do not include magnetic fields or the potential effects of dark matter annihilation when calculating the stability of haloes against gravitational collapse. It is likely that the gas in primordial halos was substantially magnetised, because any pre-existing field was easily amplified to dynamically significant levels by the small-scale turbulent dynamo. This process converts parts of the kinetic energy of the halo gas into magnetic energy \citep{schleicher08, schleicher10, sur10, schober12}.
Also, if dark matter particles are self-annihilating, then the additional heat generated by this process could influence the star formation process \citep[e.g.][]{freese08, iocco08, spolyar09, schleicher09a, schleicher09b, smith12}. Both effects increase the minimum mass for collapse.
Moreover, our star formation criteria are just a first approximation. The critical mass threshold alone might not be sufficient to decide whether a dark matter halo can collapse to form primordial stars (M. Sasaki, priv. comm.). Similar holds for the transition from Pop~III to Pop~I/II star formation. A more sophisticated approach in the future should include a more detailed treatment of metal mixing and a better description of ionising radiation. 

We also mention that our approach does not account for the possibility that Pop~III survivors might be polluted with metals after they have formed, either within the original halo or later on, once they have become part of the Galaxy \cite[see, e.g.,][]{frebel09, johnson14}. The recent detection of an extremely iron-poor Pop II star \citep{ketal14} suggests that pollution does not significantly affect the observed metallicities of all of the currently known EMP stars, but the fraction that are significantly affected has yet to be properly quantified. Most ancient open clusters show chemical homogeneity \citep{hawthorn10, friel14}, which decreases the possibility that single stars are polluted. However, it is generally possible that all cluster stars have the same surface contamination. For the purposes of our current study, we assume that the effects of pollution are either negligible or can be identified and corrected for.

\subsection{Rotation}
\label{sec:rotation}
So far, our model is based on the assumption of non-rotating Pop~III stars. However, \citet{eetal08} have shown, together with other groups, that rotation can significantly influence the evolution of metal-free stars, and \citet{stacy13} illustrate the possibility of rapidly rotating Pop~III protostars.
The rotationally-induced mixing increases the mass of the He-cores at the end of the evolution, which in turn changes the final fates of the Pop~III stars. Generally, rotation leads to longer lifetimes and higher metal yields. Hence, we expect stronger feedback effects via metal pollution, LW radiation, and ionising photons. Furthermore, the mass ranges of the final fates of primordial stars slightly change. To study the effect of rotation on the number of Pop~III survivors, we assume all Pop~III stars to be strongly rotating, with the corresponding stellar lifetimes and ejecta masses \citep{eetal08}. For the mass ranges of primordial stars 
that end 
as black holes, we use the He-core masses at the pre-SN stage, interpolate between them, and compare to the results of \citet{hw02}. This yields BH remnants for initial masses of Pop~III stars in the range $\sim 25-75M_\odot$ and above $240 M_\odot$. Assuming the same production rate of ionising photons, the SFE has to be smaller to compensate for the longer lifetimes of the stars. The best-fitting value of $\eta _{*,{\rm rot}} =0.004$ yields an optical depth of $\tau = 0.0957$. The mass accreted onto BHs is still below the limit given by the UXBG and the mean and maximal metallicity are slightly higher compared to the values in Fig.\;\ref{fig:zZ}, but are still in compliance with the observed DLAs. The corresponding numbers of expected survivors can be seen in Fig.\;\ref{fig:parameters}.
A proper treatment of rotation might also account for the different stellar spectra of rotating stars and include a distribution of rotation velocities, but from this first analysis we already see that the differences in the number of survivors are not huge. Consequently, the question of whether Pop~III stars are rapidly rotating or not, has no significant influence of the results of this study.

\subsection{IMF Dependence}
\label{sec:IMFdep}
The actual shape of the primordial IMF is the largest unknown in our model. While we explicitly vary and try to constrain the lower mass end in this paper, the actual slope and the high-mass end are still uncertain. Here, we wish to study the sensitivity of our results to changes in the IMF high-mass end and its slope. Whereas a logarithmically flat IMF seems to agree best with current simulations \citep{getal11b,dgck13,hetal14}, we also want to test the assumption of a Salpeter IMF with a slope of $\alpha = -1.35$ \citep{s55}. To match the optical depth criterion, we have to use a SFE of $\eta _* = 0.02$ in this case. The results in Fig.\;\ref{fig:parameters} show clearly that a Salpeter IMF yields a higher number of Pop~III survivors, because almost no Pop~III stars might be massive enough to produce sufficiently many ionising photons to reproduce the optical depth. We note, however, that the numerical simulations do not favour such a bottom-heavy IMF and a higher number of Pop~III 
survivors, casued by a steeper IMF, might even strengthen our conclusion.

The high-mass end of the IMF limits the amount of gas that is left for low-mass stars and possible survivors. The associated star formation efficiencies are $\eta_*(60M_\odot)=0.03$, $\eta_*(140M_\odot)=10^{-3}$, and $\eta_*(170M_\odot)=10^{-4}$ to match the optical depth constraint. Especially for masses $<100 M_\odot$, the number of survivors depends crucially on this parameter. Specifically, $M_\mathrm{max}<100 M_\odot$ might yield more possible survivors, thus strengthening our argument, whereas $100 M_\odot<M_\mathrm{max}<170 M_\odot$ does not significantly reduce the number of survivors. This simple analysis shows that the value of $M_\mathrm{max} = 100 M_\odot$ for our fiducial model is not only in good agreement with simulations, but also yields a reasonable lower limit on the number of expected Pop~III survivors and hence supports our final conclusion.

\subsection{Reionisation History}
\label{sec:ReionHist}
\begin{figure}
\centering
\includegraphics[angle=-90]{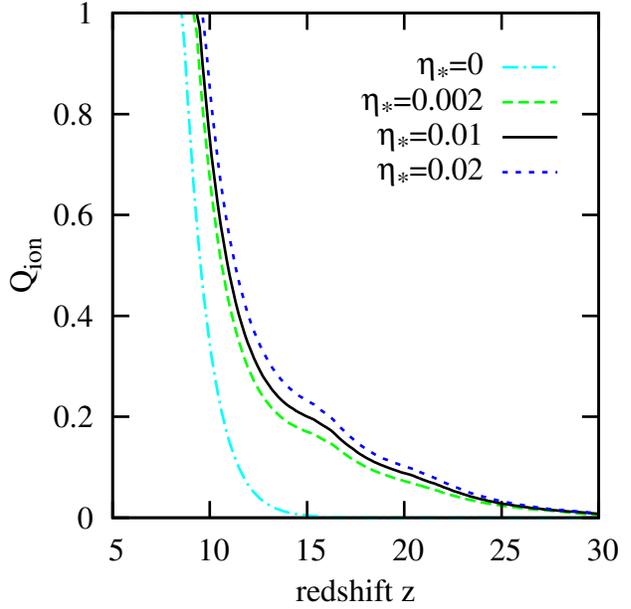}
\caption{Volume filling fraction of the ionised IGM as a function of redshift. Values for the standard case ($f_{\rm esc,III}=0.7$ and $f_{\rm esc}=0.3$), whereas $\eta _* =0$ represents the Pop I/II-only model for comparison. In all models, reionisation occurs between $z=8-10$.}
\label{fig:Qion}
\end{figure}

The escape fraction of ionising radiation is very uncertain and might vary with redshift and halo mass. Here, we study the dependence of the escape fraction on our model predictions. Therefore, we vary the fiducial escape fractions of $f_{{\rm esc,III}}=0.7$ and $f_{\rm esc}=0.3$ to $f_{\rm esc,III}=0.6$ and $f_{\rm esc}=0.2$, and to $f_{\rm esc,III}=0.8$ and $f_{\rm esc}=0.4$, adjusting the SFE accordingly ($\eta _* = 0.025$ for low escape fractions and $\eta _* = 0.002$ for high escape fractions) to reproduce the optical depth and compare the predicted number of survivors for each case. The reionisation history can be seen in Fig.\;\ref{fig:Qion} for our standard model, and for different star formation efficiencies.
Even for these different escape fractions, the expected numbers of Pop~III survivors do not change significantly. Consequently, the predictions by our model are not very sensitive on the exact reionisation history.

\section{Summary and Conclusions}
We have used a semi-analytical model to determine the Pop III star formation history in a Milky Way-like halo. This model, which was based on a Monte Carlo sampling of merger trees, contained radiative and chemical feedback to self-consistently suppress Pop~III star formation at smaller redshifts. We were able to reproduce a suite of independent observations like the optical depth to Thomson scattering and the metal enrichment history.
Primordial stars with masses below $\sim 0.8 M_\odot$ might have survived until today and should be observable in large surveys. Comparing the expected number of Pop~III survivors in our model to estimates of current survey sizes enables us to constrain the lower mass IMF limit to $0.65 M_\odot$ with a confidence of $95\%$.
For non-detections to exclude any Pop~III survivor, and hence set $M_\mathrm{min} \geq 0.8 M_\odot$, the future surveys need to include at least $\sim 2 \times 10^7$ halo stars.
In order to draw the same conclusions from observations of stars in the Milky Way bulge, we have to observe $\sim 10^3$ times more stars. This comes from the fact that there are generally more stars in the bulge and the absolute number of Pop~III survivors is smaller there. In addition, the determination of stellar metallicities is hindered by dust distinction and consequently the current sample sizes for bulge stars are smaller, compared to those of the Milky Way halo.

By modelling the history of Pop~III star formation self-consistently, we are also able to determine the most important feedback mechanisms and related quantities. For most of the time, the critical mass threshold for H$_2$ cooling is the dominant suppression mechanism for primordial star formation and only at smaller redshifts the radiative and chemical feedback becomes important. We find that the Pop~III star formation rate peaks at $z\sim 20$ and that each successfully Pop~III forming halo has an average stellar mass content of $\sim 100M_\odot$. However, this stellar mass content per halo has a huge scatter between $25-500 M_\odot$, reflecting the stochastic nature of star formation. Also the multiplicity of Pop~III systems shows a large scatter with a clear trend to systems with multiple stars.

The presented model and statistical analysis is based on several assumptions and simplifications. 
However, we have shown that the conclusions are very robust with respect to changes in fundamental model parameters. Moreover, most assumptions in our model tend to underestimate the number of Pop~III survivors and therefore even strengthen our final conclusion.
This model can be improved by including the proper spatial distribution of the haloes, or a self-consistent transition from Pop~III to subsequent star formation modes. The current statistical model assumes that all stars have the same observability and that they are selected randomly. To improve this model, one could think of a weighting function, taking into account different degrees of observability together with selection criteria for follow-up spectroscopic surveys.

In future work, we will compare our analytical model to three-dimensional simulations to obtain better estimates for the spatial distribution, which is important for a proper feedback model at higher redshifts, for the planning of further surveys in the present-day halo. Moreover, a similar study with the focus on satellite galaxies and ultra-faint dwarfs could reveal interesting additional results and better constraints, in order to provide a well-considered observation plan for upcoming surveys.

Besides a fundamental insight in the primordial star formation history and an estimation for the low mass end of the Pop~III IMF, we also provide a target sample size for upcoming surveys. Hitherto, the prime target of stellar archaeology was to identify individual intriguing, record-setting EMP stars; now, we are entering a novel phase of discovery, where statistically representative samples with well-controlled selection and incompleteness biases will become key goals. Put differently: Even the absence of record-setting stars provides powerful constraints, provided that the underlying surveys are fairly sampling the metal-poor tail of the Galactic stellar system.

\subsection*{Acknowledgements}
TH thanks Jeremy Ritter for his hospitality and fruitful discussions during the stay in Austin. Moreover, we would like to thank Marta Volonteri, Ken Freeman, Norbert Christlieb, Paul Clark, Mei Sasaki, Anna Frebel, Brendan Griffen, Greg Dooley, Alex Ji, and the anonymous referee for helpful comments and contributions. TH, RSK, and SCOG are supported from the European Research Council under the European Community's Seventh Framework Programme (FP7/2007-2013) via the ERC Advanced Grant `STARLIGHT: Formation of the First Stars' (project number 339177). The simulations described in this paper were performed on the {\em kolob} cluster at the University of Heidelberg, which is funded in part by the DFG via Emmy-Noether grant BA 3706, and via a Frontier grant of Heidelberg University, sponsored by the German Excellence Initiative as well as the Baden-W\"urttemberg Foundation. Moreover, TH acknowledges funding from the European Research Council under the European Community’s Seventh Framework Programme (
FP7/2007-2013 Grant Agreement 
no. 614199, project `BLACK'). VB 
acknowledges support from NSF grants AST-1009928 and AST-1413501. SCOG acknowledges support from the DFG via SFB 881, `The Milky Way System' (sub-projects B1, B2 and B8). The authors have employed the publicly available {\sc galform} and {\sc camb} codes.

\bibliographystyle{mn2e}

\begin{thebibliography}{}

\bibitem[\protect\citeauthoryear{{Abel}, {Bryan} \& {Norman}}{{Abel}
  et~al.}{2002}]{abn02}
{Abel} T.,  {Bryan} G.~L.,    {Norman} M.~L.,  2002, Science, 295, 93

\bibitem[\protect\citeauthoryear{{Alvarez}, {Wise} \& {Abel}}{{Alvarez}
  et~al.}{2009}]{awa09}
{Alvarez} M.~A.,  {Wise} J.~H.,    {Abel} T.,  2009, \apj, 701, L133

\bibitem[\protect\citeauthoryear{{Aoki}, {Tominaga}, {Beers}, {Honda} \&
  {Lee}}{{Aoki} et~al.}{2014}]{aoki14}
{Aoki} W.,  {Tominaga} N.,  {Beers} T.~C.,  {Honda} S.,    {Lee} Y.~S.,  2014,
  Science, 345, 912

\bibitem[\protect\citeauthoryear{{Barkana} \& {Loeb}}{{Barkana} \&
  {Loeb}}{2007}]{barkana07}
{Barkana} R.,  {Loeb} A.,  2007, Rep. Prog. Phys., 70, 627

\bibitem[\protect\citeauthoryear{{Beers} \& {Christlieb}}{{Beers} \&
  {Christlieb}}{2005}]{bc05}
{Beers} T.~C.,  {Christlieb} N.,  2005, \araa, 43, 531

\bibitem[\protect\citeauthoryear{{Bland-Hawthorn} \&
  {Maloney}}{{Bland-Hawthorn} \& {Maloney}}{1999}]{bland99}
{Bland-Hawthorn} J.,  {Maloney} P.~R.,  1999, \apj, 510, L33

\bibitem[\protect\citeauthoryear{{Bland-Hawthorn} \&
  {Peebles}}{{Bland-Hawthorn} \& {Peebles}}{2006}]{bhp06}
{Bland-Hawthorn} J.,  {Peebles} P.~J.~E.,  2006, Science, 313, 311

\bibitem[\protect\citeauthoryear{{Bland-Hawthorn}, {Krumholz} \&
  {Freeman}}{{Bland-Hawthorn} et~al.}{2010}]{hawthorn10}
{Bland-Hawthorn} J.,  {Krumholz} M.~R., {Freeman} K.,  2010, \apj, 713, 166

\bibitem[\protect\citeauthoryear{{Blum}}{{Blum}}{1995}]{b95}
{Blum} R.~D.,  1995, \apj, 444, L89

\bibitem[\protect\citeauthoryear{{Bond}, {Cole}, {Efstathiou} \&
  {Kaiser}}{{Bond} et~al.}{1991}]{bcek91}
{Bond} J.~R.,  {Cole} S.,  {Efstathiou} G.,    {Kaiser} N.,  1991, \apj, 379,
  440

\bibitem[\protect\citeauthoryear{{Bromm}}{{Bromm}}{2013}]{bromm13}
{Bromm} V.,  2013, Rep. Prog. Phys., 76, 112901

\bibitem[\protect\citeauthoryear{{Bromm}, {Coppi} \& {Larson}}{{Bromm}
  et~al.}{2002}]{bcl02}
{Bromm} V.,  {Coppi} P.~S.,    {Larson} R.~B.,  2002, \apj, 564, 23

\bibitem[\protect\citeauthoryear{{Brook}, {Kawata}, {Scannapieco}, {Martel} \&
  {Gibson}}{{Brook} et~al.}{2007}]{betal07}
{Brook} C.~B.,  {Kawata} D.,  {Scannapieco} E.,  {Martel} H.,    {Gibson}
  B.~K.,  2007, \apj, 661, 10

\bibitem[\protect\citeauthoryear{{Caffau}, {Bonifacio}, {Fran{\c c}ois}, 
{Sbordone}, {Monaco} \& et al.}{{Caffau} et~al.}{2011}]{caffau11}
{Caffau} E., {Bonifacio} P., {Fran{\c c}ois} P.,
{Sbordone} L., {Monaco} L., et al., 2011, \nat, 477, 67
  
\bibitem[\protect\citeauthoryear{{Caffau}, {Bonifacio}, {Sbordone},
{Fran{\c c}ois}, {Monaco} \& et al.}{{Caffau} et~al.}{2013}]{caffau13}
{Caffau} E., {Bonifacio} P., {Sbordone} L., {Fran{\c c}ois} P.,
{Monaco} L., et al., 2013, \aap, 560, A71
  
\bibitem[\protect\citeauthoryear{{Campisi}, {Maio}, {Salvaterra} \&
  {Ciardi}}{{Campisi} et~al.}{2011}]{cmsc11}
{Campisi} M.~A.,  {Maio} U.,  {Salvaterra} R.,    {Ciardi} B.,  2011, \mnras,
  416, 2760

\bibitem[\protect\citeauthoryear{{Christlieb}, {Sch{\"o}rck}, {Frebel},
  {Beers}, {Wisotzki} \& {Reimers}}{{Christlieb} et~al.}{2008}]{csfbw08}
{Christlieb} N.,  {Sch{\"o}rck} T.,  {Frebel} A.,  {Beers} T.~C.,  {Wisotzki}
  L.,    {Reimers} D.,  2008, \aap, 484, 721

\bibitem[\protect\citeauthoryear{{Clark}, {Glover}, {Klessen} \&
  {Bromm}}{{Clark} et~al.}{2011a}]{cgkb11}
{Clark} P.~C.,  {Glover} S.~C.~O.,  {Klessen} R.~S.,    {Bromm} V.,  2011a, \apj, 727, 110

\bibitem[\protect\citeauthoryear{{Clark}, {Glover}, {Smith}, {Greif}, {Klessen}
  \& {Bromm}}{{Clark} et~al.}{2011b}]{clark11}
{Clark} P.~C.,  {Glover} S.~C.~O.,  {Smith} R.~J.,  {Greif} T.~H.,  {Klessen}
  R.~S.,    {Bromm} V.,  2011b, Science, 331, 1040

\bibitem[\protect\citeauthoryear{{Cole}, {Lacey}, {Baugh} \& {Frenk}}{{Cole}
  et~al.}{2000}]{clbf00}
{Cole} S.,  {Lacey} C.~G.,  {Baugh} C.~M.,    {Frenk} C.~S.,  2000, \mnras,
  319, 168

\bibitem[\protect\citeauthoryear{{de Bennassuti}, {Schneider}, {Valiante} \& {Salvadori}}
{{de Bennassuti} et~al.}{2014}]{debennassuti14}
{de Bennassuti} M., {Schneider} R., {Valiante} R., {Salvadori} S.,  2014, \mnras, 445, 3039
  
\bibitem[\protect\citeauthoryear{{Diemand}, {Madau} \& {Moore}}{{Diemand}
  et~al.}{2005}]{dmm05}
{Diemand} J.,  {Madau} P.,    {Moore} B.,  2005, \mnras, 364, 367

\bibitem[\protect\citeauthoryear{{Dopcke}, {Glover}, {Clark} \&
  {Klessen}}{{Dopcke} et~al.}{2013}]{dgck13}
{Dopcke} G.,  {Glover} S.~C.~O.,  {Clark} P.~C.,    {Klessen} R.~S.,  2013,
  \apj, 766, 103

\bibitem[\protect\citeauthoryear{{Draine}}{{Draine}}{2011}]{d11}
{Draine} B.~T.,  2011, {Physics of the Interstellar and Intergalactic Medium}.
Princeton Univ. Press, Princeton

\bibitem[\protect\citeauthoryear{{Ekstr{\"o}m}, {Meynet}, {Chiappini},
  {Hirschi} \& {Maeder}}{{Ekstr{\"o}m} et~al.}{2008}]{eetal08}
{Ekstr{\"o}m} S.,  {Meynet} G.,  {Chiappini} C.,  {Hirschi} R.,    {Maeder} A.,
   2008, \aap, 489, 685

\bibitem[\protect\citeauthoryear{{Frebel}}{{Frebel}}{2010}]{f10}
{Frebel} A.,  2010, Astron. Nachr., 331, 474

\bibitem[\protect\citeauthoryear{{Frebel}, {Johnson} \& {Bromm}}{{Frebel}
  et~al.}{2009}]{frebel09}
{Frebel} A.,  {Johnson} J.~L.,    {Bromm} V.,  2009, \mnras, 392, L50

\bibitem[\protect\citeauthoryear{{Freeman} \& {Bland-Hawthorn}}{{Freeman} \&
  {Bland-Hawthorn}}{2002}]{fbh02}
{Freeman} K.,  {Bland-Hawthorn} J.,  2002, \araa, 40, 487

\bibitem[\protect\citeauthoryear{{Freese}, {Bodenheimer}, {Spolyar} \& {Gondolo}}{{Freese}
  et~al.}{2008}]{freese08}
{Freese} K.,  {Bodenheimer} P., {Spolyar} D., {Gondolo} P., 2008, \apj, 685, L101

\bibitem[\protect\citeauthoryear{{Friel}, {Donati}, {Bragaglia}, {Jacobson},
{Magrini}, {Prisinzano}, {Randich} \& et~al.}
{{Friel} et~al.}{2014}]{friel14}
{Friel} E.~D., {Donati} P., {Bragaglia} A., {Jacobson} H.~R.,
{Magrini} L., {Prisinzano} L., {Randich} S., et~al. 2014, \aap, 563, A117

\bibitem[\protect\citeauthoryear{{Garc{\'{\i}}a P{\'e}rez}, {Cunha},
  {Shetrone}, {Majewski}, {Johnson}, {Smith}, {Schiavon} \& et.
  al}{{Garc{\'{\i}}a P{\'e}rez} et~al.}{2013}]{gpetal2013}
{Garc{\'{\i}}a P{\'e}rez} A.~E.,  {Cunha} K.,  {Shetrone} M.,  {Majewski}
  S.~R.,  {Johnson} J.~A.,  {Smith} V.~V.,  {Schiavon} R.~P.,    et. al 2013,
  \apj, 767, L9

\bibitem[\protect\citeauthoryear{{Glover} \& {Brand}}{{Glover} \&
  {Brand}}{2003}]{glover03}
{Glover} S.~C.~O., {Brand} P.~W.~J.~L., 2003, \mnras, 340, 210
  
\bibitem[\protect\citeauthoryear{{Glover}}{{Glover}}{2013}]{g13}
{Glover} S.,  2013, in {Wiklind} T.,  {Mobasher} B.,   {Bromm} V.,  eds,
  Astrophysics and Space Science Library Vol.~396 of Astrophysics and Space
  Science Library, {The First Stars}.
p.~103

\bibitem[\protect\citeauthoryear{{Greif} \& {Bromm}}{{Greif} \&
  {Bromm}}{2006}]{gb06}
{Greif} T.~H.,  {Bromm} V.,  2006, \mnras, 373, 128

\bibitem[\protect\citeauthoryear{{Greif}, {White}, {Klessen} \& {Springel}}{{Greif} et~al.}{2011a}]{getal11a}
{Greif} T.~H., {White} S.~D.~M.,{Klessen} R.~S., {Springel} V., 2011a, \apj, 736, 147

\bibitem[\protect\citeauthoryear{{Greif}, {Springel}, {White}, {Glover},
  {Clark}, {Smith}, {Klessen} \& {Bromm}}{{Greif} et~al.}{2011b}]{getal11b}
{Greif} T.~H.,  {Springel} V.,  {White} S.~D.~M.,  {Glover} S.~C.~O.,  {Clark}
  P.~C.,  {Smith} R.~J.,  {Klessen} R.~S.,    {Bromm} V.,  2011b, \apj, 737, 75

\bibitem[\protect\citeauthoryear{{Greif}, {Bromm}, {Clark}, {Glover}, {Smith},
  {Klessen}, {Yoshida} \& {Springel}}{{Greif} et~al.}{2012}]{greif12}
{Greif} T.~H.,  {Bromm} V.,  {Clark} P.~C.,  {Glover} S.~C.~O.,  {Smith} R.~J.,
   {Klessen} R.~S.,  {Yoshida} N.,    {Springel} V.,  2012, \mnras, 424, 399

\bibitem[\protect\citeauthoryear{{Greif}}{{Greif}}{2015}]{greif14}
{Greif} T.~H., 2015, Comput. Astrophys. \& Cosm., 2, 3
   
\bibitem[\protect\citeauthoryear{{Haiman}, {Thoul} \& {Loeb}}{{Haiman}
  et~al.}{1996a}]{haiman96a}
{Haiman} Z.,  {Thoul} A.~A.,    {Loeb} A.,  1996a, \apj, 464, 523
   
\bibitem[\protect\citeauthoryear{{Haiman}, {Rees} \& {Loeb}}{{Haiman}
  et~al.}{1996b}]{haiman96b}
{{Haiman} Z., {Rees} M.~J., {Loeb}, A.},  1996b, \apj, 467, 522
   
\bibitem[\protect\citeauthoryear{{Haiman}, {Abel} \& {Rees}}{{Haiman}
  et~al.}{2000}]{har00}
{Haiman} Z.,  {Abel} T.,    {Rees} M.~J.,  2000, \apj, 534, 11

\bibitem[\protect\citeauthoryear{{Hartwig}, {Clark}, {Glover}, {Klessen} \&
  {Sasaki}}{{Hartwig} et~al.}{2015}]{hcgk14}
{Hartwig} T.,  {Clark} P.~C.,  {Glover} S.~C.~O.,  {Klessen} R.~S.,    {Sasaki}
  M.,  2015, \apj, 799, 144

\bibitem[\protect\citeauthoryear{{Heger} \& {Woosley}}{{Heger} \&
  {Woosley}}{2002}]{hw02}
{Heger} A.,  {Woosley} S.~E.,  2002, \apj, 567, 532

\bibitem[\protect\citeauthoryear{{Heger} \& {Woosley}}{{Heger} \&
  {Woosley}}{2010}]{hw10}
{Heger} A.,  {Woosley} S.~E.,  2010, \apj, 724, 341

\bibitem[\protect\citeauthoryear{{Hirano}, {Hosokawa}, {Yoshida}, {Umeda},
  {Omukai}, {Chiaki} \& {Yorke}}{{Hirano} et~al.}{2014}]{hetal14}
{Hirano} S.,  {Hosokawa} T.,  {Yoshida} N.,  {Umeda} H.,  {Omukai} K.,
  {Chiaki} G.,    {Yorke} H.~W.,  2014, \apj, 781, 60

\bibitem[\protect\citeauthoryear{{Hopkins} \& {Beacom}}{{Hopkins} \&
  {Beacom}}{2006}]{hb06}
{Hopkins} A.~M.,  {Beacom} J.~F.,  2006, \apj, 651, 142

\bibitem[\protect\citeauthoryear{{Hosokawa}, {Yoshida}, {Omukai} \&
  {Yorke}}{{Hosokawa} et~al.}{2012}]{hyoy12}
{Hosokawa} T.,  {Yoshida} N.,  {Omukai} K.,    {Yorke} H.~W.,  2012, \apj, 760,
  L37

\bibitem[\protect\citeauthoryear{{Iocco}}{{Iocco}}{2008}]{iocco08}
{Iocco} F.,  2008, \apj, 677, L1

\bibitem[\protect\citeauthoryear{{Islam}, {Taylor} \&
  {Silk}}{{Islam} et~al.}{2003}]{islam03}
{Islam} R.~R., {Taylor} J.~E., {Silk} J., 2003, \mnras, 340, 647

\bibitem[\protect\citeauthoryear{{Islam}, {Taylor} \&
  {Silk}}{{Islam} et~al.}{2004a}]{islam04a}
{Islam} R.~R., {Taylor} J.~E., {Silk} J., 2004a, \mnras, 354, 427
  
\bibitem[\protect\citeauthoryear{{Islam}, {Taylor} \&
  {Silk}}{{Islam} et~al.}{2004b}]{islam04b}
{Islam} R.~R., {Taylor} J.~E., {Silk} J., 2004b, \mnras, 354, 443

\bibitem[\protect\citeauthoryear{{Jeon}, {Pawlik}, {Bromm} \&
  {Milosavljevi{\'c}}}{{Jeon} et~al.}{2014}]{jpbm14}
{Jeon} M.,  {Pawlik} A.~H.,  {Bromm} V.,    {Milosavljevi{\'c}} M.,  2014,
  \mnras, 440, 3778

\bibitem[\protect\citeauthoryear{{Jiang} \& {van den Bosch}}{{Jiang} \& {van
  den Bosch}}{2014}]{jb14}
{Jiang} F.,  {van den Bosch} F.~C.,  2014, \mnras, 440, 193

\bibitem[\protect\citeauthoryear{{Johnson} \& {Bromm}}{{Johnson} \&
  {Bromm}}{2006}]{johnson06}
{Johnson} J.~L.,  {Bromm} V.,  2006, \mnras, 366, 247

\bibitem[\protect\citeauthoryear{{Johnson}, {Greif} \& {Bromm}}{{Johnson}
  et~al.}{2008}]{jgb08}
{Johnson} J.~L.,  {Greif} T.~H.,    {Bromm} V.,  2008, \mnras, 388, 26

\bibitem[\protect\citeauthoryear{{Johnson}}{{Johnson}}{2014}]{johnson14}
{Johnson} J.~L., 2014, arXiv:1411.4189

\bibitem[\protect\citeauthoryear{{Karlsson}, {Bromm} \&
  {Bland-Hawthorn}}{{Karlsson} et~al.}{2013}]{kbbh13}
{Karlsson} T.,  {Bromm} V.,    {Bland-Hawthorn} J.,  2013, Rev. Mod. Phys., 85,
  809

\bibitem[\protect\citeauthoryear{{Karlsson}, {Johnson} \& {Bromm}}{{Karlsson}
  et~al.}{2008}]{kjb08}
{Karlsson} T.,  {Johnson} J.~L.,    {Bromm} V.,  2008, \apj, 679, 6

\bibitem[\protect\citeauthoryear{{Keller}, {Bessell}, {Frebel}, {Casey},
  {Asplund}, {Jacobson}, {Lind}, {Norris}, {Yong}, {Heger}, {Magic}, {da
  Costa}, {Schmidt} \& {Tisserand}}{{Keller} et~al.}{2014}]{ketal14}
{Keller} S.~C.,  {Bessell} M.~S.,  {Frebel} A.,  {Casey} A.~R.,  {Asplund} M.,
  {Jacobson} H.~R.,  {Lind} K.,  {Norris} J.~E.,  {Yong} D.,  {Heger} A.,
  {Magic} Z.,  {da Costa} G.~S.,  {Schmidt} B.~P.,    {Tisserand} P.,  2014,
  \nat, 506, 463

\bibitem[\protect\citeauthoryear{{Kitayama}, {Yoshida}, {Susa} \&
  {Umemura}}{{Kitayama} et~al.}{2004}]{kitayama04}
{Kitayama} T., {Yoshida} N., {Susa} H., {Umemura} M.,  2004, \apj,
  613, 631  
  
\bibitem[\protect\citeauthoryear{{Kroupa}}{{Kroupa}}{2001}]{k01}
{Kroupa} P.,  2001, \mnras, 322, 231

\bibitem[\protect\citeauthoryear{{Kulkarni}, {Hennawi}, {Rollinde} \&
  {Vangioni}}{{Kulkarni} et~al.}{2014}]{khrv14}
{Kulkarni} G.,  {Hennawi} J.~F.,  {Rollinde} E.,    {Vangioni} E.,  2014, \apj,
  787, 64

\bibitem[\protect\citeauthoryear{{Kulkarni}, {Rollinde}, {Hennawi} \&
  {Vangioni}}{{Kulkarni} et~al.}{2013}]{krhv13}
{Kulkarni} G.,  {Rollinde} E.,  {Hennawi} J.~F.,    {Vangioni} E.,  2013, \apj,
  772, 93

\bibitem[\protect\citeauthoryear{{Lacey} \& {Cole}}{{Lacey} \&
  {Cole}}{1993}]{lc93}
{Lacey} C.,  {Cole} S.,  1993, \mnras, 262, 627

\bibitem[\protect\citeauthoryear{{Lewis}, {Challinor} \& {Lasenby}}{{Lewis}
  et~al.}{2000}]{lcl00}
{Lewis} A.,  {Challinor} A.,    {Lasenby} A.,  2000, \apj, 538, 473

\bibitem[\protect\citeauthoryear{{Li}}{{Li}}{2008}]{l08}
{Li} L.-X.,  2008, \mnras, 388, 1487

\bibitem[\protect\citeauthoryear{{Loeb}}{{Loeb}}{2010}]{l10}
{Loeb} A.,  2010, {How Did the First Stars and Galaxies Form?}.
Princeton Univ. Press, Princeton

\bibitem[\protect\citeauthoryear{{Loeb} \& {Furlanetto}}{{Loeb} \&
  {Furlanetto}}{2013}]{loeb13}
{Loeb} A.,  {Furlanetto} S.~R.,  2013, {The First Galaxies in the Universe}.
Princeton Univ. Press, Princeton

\bibitem[\protect\citeauthoryear{{Machacek}, {Bryan} \& {Abel}}{{Machacek}
  et~al.}{2001}]{mba01}
{Machacek} M.~E.,  {Bryan} G.~L.,    {Abel} T.,  2001, \apj, 548, 509

\bibitem[\protect\citeauthoryear{{Machacek}, {Bryan} \& {Abel}}{{Machacek}
  et~al.}{2003}]{machacek03}
{Machacek} M.~E., {Bryan} G.~L., {Abel}, T.,  2003, \mnras, 338, 273

\bibitem[\protect\citeauthoryear{{Madau} \& {Dickinson}}{{Madau} \&
  {Dickinson}}{2014}]{md14}
{Madau} P.,  {Dickinson} M.,  2014, \araa, 52, 415

\bibitem[\protect\citeauthoryear{{Maio}, {Koopmans} \&
  {Ciardi}}{{Maio} et~al.}{2011}]{maio11}
{Maio} U.,  {Koopmans} L.~V.~E., {Ciardi} B., 2011, \mnras, 412, L40

\bibitem[\protect\citeauthoryear{{Mapelli}, {Ferrara} \&
  {Rea}}{{Mapelli} et~al.}{2006}]{mapelli06}
{Mapelli} M.,  {Ferrara} A., {Rea} N., 2006, \mnras, 368, 1340

\bibitem[\protect\citeauthoryear{{Marigo}, {Girardi}, {Chiosi} \&
  {Wood}}{{Marigo} et~al.}{2001}]{mgcw01}
{Marigo} P.,  {Girardi} L.,  {Chiosi} C.,    {Wood} P.~R.,  2001, \aap, 371,
  152

\bibitem[\protect\citeauthoryear{{McMillan}}{{McMillan}}{2011}]{m11}
{McMillan} P.~J.,  2011, \mnras, 414, 2446

\bibitem[\protect\citeauthoryear{{Mesler}, {Whalen}, {Smidt}, {Fryer},
  {Lloyd-Ronning} \& {Pihlstr{\"o}m}}{{Mesler} et~al.}{2014}]{mesler14}
{Mesler} R.~A.,  {Whalen} D.~J.,  {Smidt} J.,  {Fryer} C.~L.,  {Lloyd-Ronning}
  N.~M.,    {Pihlstr{\"o}m} Y.~M.,  2014, \apj, 787, 91

\bibitem[\protect\citeauthoryear{{Mirabel}, {Dijkstra}, {Laurent}, {Loeb} \&
  {Pritchard}}{{Mirabel} et~al.}{2011}]{mirabel11}
{Mirabel} I.~F.,  {Dijkstra} M.,  {Laurent} P.,  {Loeb} A.,    {Pritchard}
  J.~R.,  2011, \aap, 528, A149

\bibitem[\protect\citeauthoryear{{Moretti}, {Vattakunnel}, {Tozzi},
  {Salvaterra}, {Severgnini}, {Fugazza}, {Haardt} \& {Gilli}}{{Moretti}
  et~al.}{2012}]{metal12}
{Moretti} A.,  {Vattakunnel} S.,  {Tozzi} P.,  {Salvaterra} R.,  {Severgnini}
  P.,  {Fugazza} D.,  {Haardt} F.,    {Gilli} R.,  2012, \aap, 548, A87

\bibitem[\protect\citeauthoryear{{Nakamura} \& {Umemura}}{{Nakamura} \&
  {Umemura}}{2002}]{nu02}
{Nakamura} F.,  {Umemura} M.,  2002, \apj, 569, 549

\bibitem[\protect\citeauthoryear{{Ness}, {Freeman}, {Athanassoula},
  {Wylie-de-Boer}, {Bland-Hawthorn}, {Asplund}, {Lewis}, {Yong}, {Lane} \&
  {Kiss}}{{Ness} et~al.}{2013}]{ness13}
{Ness} M.,  {Freeman} K.,  {Athanassoula} E.,  {Wylie-de-Boer} E.,
  {Bland-Hawthorn} J.,  {Asplund} M.,  {Lewis} G.~F.,  {Yong} D.,  {Lane}
  R.~R.,    {Kiss} L.~L.,  2013, \mnras, 430, 836

\bibitem[\protect\citeauthoryear{{O'Shea} \& {Norman}}{{O'Shea} \&
  {Norman}}{2008}]{on08}
{O'Shea} B.~W.,  {Norman} M.~L.,  2008, \apj, 673, 14

\bibitem[\protect\citeauthoryear{{Oh}}{{Oh}}{2001}]{oh01}
{Oh} S.~P., 2001, \apj, 553, 499
  
\bibitem[\protect\citeauthoryear{{Omukai} \& {Palla}}{{Omukai} \&
  {Palla}}{2001}]{op01}
{Omukai} K.,  {Palla} F.,  2001, \apj, 561, L55

\bibitem[\protect\citeauthoryear{{Omukai} \& {Palla}}{{Omukai} \&
  {Palla}}{2003}]{op03}
{Omukai} K.,  {Palla} F.,  2003, \apj, 589, 677

\bibitem[\protect\citeauthoryear{{Paardekooper}, {Khochfar} \& {Dalla
  Vecchia}}{{Paardekooper} et~al.}{2013}]{pkd13}
{Paardekooper} J.-P.,  {Khochfar} S.,    {Dalla Vecchia} C.,  2013, \mnras,
  429, L94

\bibitem[\protect\citeauthoryear{{Pallottini}, {Ferrara}, {Gallerani}, {Salvadori} \& {D'Odorico}}{{Pallottini} et~al.}{2014}]{pallottini14}
{Pallottini}, A., {Ferrara}, A., {Gallerani}, S., {Salvadori}, S., 
	{D'Odorico}, V.,  2014, \mnras,
  440, 2498
  
\bibitem[\protect\citeauthoryear{{Parkinson}, {Cole} \& {Helly}}{{Parkinson}
  et~al.}{2008}]{pch08}
{Parkinson} H.,  {Cole} S.,    {Helly} J.,  2008, \mnras, 383, 557

\bibitem[\protect\citeauthoryear{{Planck Collaboration}}{{Planck
  Collaboration}}{2014}]{planck13}
{Planck Collaboration} 2014, A\&A, 571, 16

\bibitem[\protect\citeauthoryear{{Power}, {Wynn}, {Combet} \&
  {Wilkinson}}{{Power} et~al.}{2009}]{pwcw09}
{Power} C.,  {Wynn} G.~A.,  {Combet} C.,    {Wilkinson} M.~I.,  2009, \mnras,
  395, 1146

\bibitem[\protect\citeauthoryear{{Press} \& {Schechter}}{{Press} \&
  {Schechter}}{1974}]{ps74}
{Press} W.~H.,  {Schechter} P.,  1974, \apj, 187, 425

\bibitem[\protect\citeauthoryear{{Rafelski}, {Wolfe}, {Prochaska}, {Neeleman}
  \& {Mendez}}{{Rafelski} et~al.}{2012}]{retal12}
{Rafelski} M.,  {Wolfe} A.~M.,  {Prochaska} J.~X.,  {Neeleman} M.,    {Mendez}
  A.~J.,  2012, \apj, 755, 89

\bibitem[\protect\citeauthoryear{{Rees}}{{Rees}}{1976}]{rees76}
{Rees} M.~J.,  1976, \mnras, 176, 483

\bibitem[\protect\citeauthoryear{{Ritter}, {Sluder}, {Safranek-Shrader},
  {Milosavljevic} \& {Bromm}}{{Ritter} et~al.}{2014}]{ritter14}
{Ritter} J.~S.,  {Sluder} A.,  {Safranek-Shrader} C.,  {Milosavljevic} M.,
  {Bromm} V.,  2014, arXiv:1408.0319

\bibitem[\protect\citeauthoryear{{Robertson}, {Furlanetto}, {Schneider},
  {Charlot}, {Ellis}, {Stark}, {McLure}, {Dunlop}, {Koekemoer}, {Schenker},
  {Ouchi}, {Ono}, {Curtis-Lake}, {Rogers}, {Bowler} \& {Cirasuolo}}{{Robertson}
  et~al.}{2013}]{retal13}
{Robertson} B.~E.,  {Furlanetto} S.~R.,  {Schneider} E.,  {Charlot} S.,
  {Ellis} R.~S.,  {Stark} D.~P.,  {McLure} R.~J.,  {Dunlop} J.~S.,  {Koekemoer}
  A.,  {Schenker} M.~A.,  {Ouchi} M.,  {Ono} Y.,  {Curtis-Lake} E.,  {Rogers}
  A.~B.,  {Bowler} R.~A.~A.,    {Cirasuolo} M.,  2013, \apj, 768, 71

\bibitem[\protect\citeauthoryear{{Salpeter}}{{Salpeter}}{1955}]{s55}
{Salpeter} E.~E.,  1955, \apj, 121, 161

\bibitem[\protect\citeauthoryear{{Salvaterra}, {Haardt}, {Volonteri} \&
  {Moretti}}{{Salvaterra} et~al.}{2012}]{shvm12}
{Salvaterra} R.,  {Haardt} F.,  {Volonteri} M.,    {Moretti} A.,  2012, \aap,
  545, L6

\bibitem[\protect\citeauthoryear{{Salvadori}, {Schneider} \& {Ferrara}}
{{Salvadori} et~al.}{2007}]{salvadori07}
{Salvadori} S., {Schneider} R., {Ferrara} A., 2007, \mnras, 381, 647
  
  \bibitem[\protect\citeauthoryear{{Salvadori}, {Ferrara}, {Schneider},
  {Scannapieco} \& {Kawata}}{{Salvadori} et~al.}{2010}]{salvadori10}
{Salvadori} S., {Ferrara} A., {Schneider} R., {Scannapieco} E., {Kawata} D. 2010, \mnras,
  401, L5
  
\bibitem[\protect\citeauthoryear{{Salvadori} \& {Ferrara}}
{{Salvadori} \& {Ferrara}}{2012}]{salvadori12}
{Salvadori} S., {Ferrara} A., 2012, \mnras, 421, L29
  
\bibitem[\protect\citeauthoryear{{Salvadori}, {Tolstoy}, {Ferrara} \& {Zaroubi}}
{{Salvadori} et~al.}{2014}]{salvadori14}
{Salvadori} S., {Tolstoy} E., {Ferrara} A., {Zaroubi} S., 2014, \mnras, 421, L29
  
  \bibitem[\protect\citeauthoryear{{Sasaki}, {Clark}, {Springel}, {Klessen} \&
  {Glover}}{{Sasaki} et~al.}{2014}]{sasaki14}
{{Sasaki} M., {Clark} P.~C., {Springel} V., {Klessen} R.~S., {Glover} S.~C.~O.}, 2014, \mnras, 442, 1942
  
\bibitem[\protect\citeauthoryear{{Scannapieco}, {Kawata}, {Brook}, {Schneider},
  {Ferrara} \& {Gibson}}{{Scannapieco} et~al.}{2006}]{setal06}
{Scannapieco} E.,  {Kawata} D.,  {Brook} C.~B.,  {Schneider} R.,  {Ferrara} A.,
     {Gibson} B.~K.,  2006, \apj, 653, 285

\bibitem[\protect\citeauthoryear{{Schaerer}}{{Schaerer}}{2002}]{s02}
{Schaerer} D.,  2002, \aap, 382, 28

\bibitem[\protect\citeauthoryear{{Schleicher}, {Banerjee} \& {Klessen}}{{Schleicher} et~al.}{2008}]{schleicher08}
{Schleicher} D.~R.~G., {Banerjee} R., {Klessen} R.~S., 2008, Phys. Rev. D, 78, 083005

\bibitem[\protect\citeauthoryear{{Schleicher}, {Banerjee} \& {Klessen}}{{Schleicher} et~al.}{2009a}]{schleicher09a}
{Schleicher} D.~R.~G., {Banerjee} R., {Klessen} R.~S., 2009a, Phys. Rev. D, 79, 043510

\bibitem[\protect\citeauthoryear{{Schleicher}, {Glover}, {Banerjee} \& {Klessen}}{{Schleicher} et~al.}{2009b}]{schleicher09b}
{Schleicher} D.~R.~G., {Glover} S.~C.~O., {Banerjee} R., {Klessen} R.~S., 2009b, Phys. Rev. D, 79, 023515

\bibitem[\protect\citeauthoryear{{Schleicher}, {Banerjee}, {Sur}, {Arshakian},
{Klessen}, {Beck} \& {Spaans}}{{Schleicher} et~al.}{2010}]{schleicher10}
{Schleicher} D.~R.~G., {Banerjee} R., {Sur} S., {Arshakian} T.~G., {Klessen} R.~S., {Beck} R.,
{Spaans} M., 2010, \aap, 522, A115

\bibitem[\protect\citeauthoryear{{Schober}, {Schleicher}, {Federrath}, {Glover},
{Klessen} \& {Banerjee}}{{Schober} et~al.}{2012}]{schober12}
Schober J., Schleicher D.~R.~G., Federrath C., Glover S.~C.~O., Klessen R.~S., {Banerjee} R., 2012, \apj, 754, 99

\bibitem[\protect\citeauthoryear{{Sedov}}{{Sedov}}{1959}]{s59}
{Sedov} L.~I.,  1959, {Similarity and Dimensional Methods in Mechanics}.
Academic Press, New York

\bibitem[\protect\citeauthoryear{Smith, Iocco, Glover, Schleicher, Klessen, Hirano 
\& Yoshida}{{Smith} et~al.}{2012}]{smith12}
Smith R.~J., Iocco F., Glover S.~C.~O., Schleicher D.~R.~G. Klessen R.~S., Hirano S., Yoshida N., 2012, \apj, 761, 154

\bibitem[\protect\citeauthoryear{Spolyar, Bodenheimer, Freese \& Gondolo}
{{Spolyar} et~al.}{2009}]{spolyar09}
Spolyar D., Bodenheimer P., Freese K., Gondolo P., 2009, \apj, 705, 1031

\bibitem[\protect\citeauthoryear{{Springel}, {White}, {Jenkins}, {Frenk},
  {Yoshida}, {Gao}, {Navarro}, {Thacker}, {Croton}, {Helly}, {Peacock}, {Cole},
  {Thomas}, {Couchman}, {Evrard}, {Colberg} \& {Pearce}}{{Springel}
  et~al.}{2005}]{setal05}
{Springel} V.,  {White} S.~D.~M.,  {Jenkins} A.,  {Frenk} C.~S.,  {Yoshida} N.,
   {Gao} L.,  {Navarro} J.,  {Thacker} R.,  {Croton} D.,  {Helly} J.,
  {Peacock} J.~A.,  {Cole} S.,  {Thomas} P.,  {Couchman} H.,  {Evrard} A.,
  {Colberg} J.,    {Pearce} F.,  2005, \nat, 435, 629

\bibitem[\protect\citeauthoryear{{Stacy}, {Greif} \& {Bromm}}{{Stacy}
  et~al.}{2010}]{sgb10}
{Stacy} A.,  {Greif} T.~H.,    {Bromm} V.,  2010, \mnras, 403, 45

\bibitem[\protect\citeauthoryear{{Stacy}, {Bromm} \& {Loeb}}{{Stacy}
  et~al.}{2011}]{stacy11}
{Stacy} A., {Bromm} V., {Loeb} A.,  2011, \apj, 730, L1

\bibitem[\protect\citeauthoryear{{Stacy}, {Greif} \& {Bromm}}{{Stacy}
  et~al.}{2012}]{sgb12}
{Stacy} A.,  {Greif} T.~H.,    {Bromm} V.,  2012, \mnras, 422, 290

\bibitem[\protect\citeauthoryear{{Stacy}, {Greif}, {Klessen}, {Bromm} \&
  {Loeb}}{{Stacy} et~al.}{2013}]{stacy13}
{Stacy} A.,  {Greif} T.~H.,  {Klessen} R.~S.,  {Bromm} V.,    {Loeb} A.,  2013,
  \mnras, 431, 1470

\bibitem[\protect\citeauthoryear{{Stacy} \& {Bromm}}{{Stacy} \&
  {Bromm}}{2014}]{sb14}
{Stacy} A.,  {Bromm} V.,  2014, \apj, 785, 73

\bibitem[\protect\citeauthoryear{Sur, Schleicher, Banerjee, Federrath \& Klessen}{{Sur} et~al.}{2010}]{sur10}
Sur S., Schleicher D.~R.~G., Banerjee R., Federrath C., Klessen R.~S.,  2010, \apj, 721, L134

\bibitem[\protect\citeauthoryear{{Taylor}}{{Taylor}}{1950}]{t50}
{Taylor} G.,  1950, Proc. Roy. Soc. A, 201, 159

\bibitem[\protect\citeauthoryear{{Trenti} \& {Stiavelli}}{{Trenti} \&
  {Stiavelli}}{2009}]{trenti09}
{Trenti} M.,  {Stiavelli} M.,  2009, \apj, 694, 879

\bibitem[\protect\citeauthoryear{{Tseliakhovich} \& {Hirata}}{{Tseliakhovich} \&
  {Hirata}}{2010}]{tseliakhovich10}
{Tseliakhovich} D.,  {Hirata} C.,  2010, Phys. Rev. D, 82, 083520

\bibitem[\protect\citeauthoryear{{Tseliakhovich}, {Barkana} \& {Hirata}}{{Tseliakhovich} et~al.}{2011}]{tseliakhovich11}
{Tseliakhovich} D., {Barkana} R., {Hirata} C.~M.,  2011, \mnras, 418, 906

\bibitem[\protect\citeauthoryear{{Tumlinson}}{{Tumlinson}}{2006}]{t06}
{Tumlinson} J.,  2006, \apj, 641, 1

\bibitem[\protect\citeauthoryear{{Tumlinson}}{{Tumlinson}}{2010a}]{t10a}
{Tumlinson} J.,  2010a, \apj, 708, 1398

\bibitem[\protect\citeauthoryear{{Tumlinson}}{{Tumlinson}}{2010b}]{t10b}
{Tumlinson} J.,  2010b, in {Whalen} D.~J.,  {Bromm} V.,   {Yoshida} N.,  eds,
  American Institute of Physics Conference Series Vol.~1294 of American
  Institute of Physics Conference Series, {The First and Second Stars: How to
  Find Them and What to Do About It (or, Three Big Ideas and Three
  Challenges)}.
pp 84--89

\bibitem[\protect\citeauthoryear{{Volonteri}}{{Volonteri}}{2012}]{v12}
{Volonteri} M.,  2012, Science, 337, 544

\bibitem[\protect\citeauthoryear{{Volonteri} \& {Bellovary}}{{Volonteri} \&
  {Bellovary}}{2012}]{volonteri12}
{Volonteri} M.,  {Bellovary} J.,  2012, Rep. Prog. Phys., 75, 124901

\bibitem[\protect\citeauthoryear{{White} \& {Rees}}{{White} \&
  {Rees}}{1978}]{wr78}
{White} S.~D.~M.,  {Rees} M.~J.,  1978, \mnras, 183, 341

\bibitem[\protect\citeauthoryear{{White} \& {Springel}}{{White} \&
  {Springel}}{2000}]{ws00}
{White} S.~D.~M.,  {Springel} V.,  2000, in {Weiss} A.,  {Abel} T.~G.,   {Hill}
  V.,  eds, The First Stars {Where Are the First Stars Now?}.
p.~327

\bibitem[\protect\citeauthoryear{{Widrow} \& {Dubinski}}{{Widrow} \&
  {Dubinski}}{2005}]{wd05}
{Widrow} L.~M.,  {Dubinski} J.,  2005, \apj, 631, 838

\bibitem[\protect\citeauthoryear{{Wise} \& {Abel}}{{Wise} \&
  {Abel}}{2007}]{wa07}
{Wise} J.~H.,  {Abel} T.,  2007, \apj, 671, 1559

\bibitem[\protect\citeauthoryear{{Yoshida}, {Abel}, {Hernquist} \&
  {Sugiyama}}{{Yoshida} et~al.}{2003}]{yahs03}
{Yoshida} N.,  {Abel} T.,  {Hernquist} L.,    {Sugiyama} N.,  2003, \apj, 592,
  645

\bibitem[\protect\citeauthoryear{{Yoshida}, {Omukai} \& {Hernquist}}{{Yoshida}
  et~al.}{2007}]{yoshida07}
{Yoshida} N.,  {Omukai} K.,    {Hernquist} L.,  2007, \apj, 667, L117

\end{thebibliography}

\bsp

\label{lastpage}

\end{document}